\documentclass[pdflatex,sn-mathphys-num]{sn-jnl}

\usepackage{graphicx}%
\usepackage{multirow}%
\usepackage{amsmath,amssymb,amsfonts}%
\usepackage{amsthm}%
\usepackage{mathrsfs}%
\usepackage[title]{appendix}%
\usepackage{xcolor}%
\usepackage{textcomp}%
\usepackage{manyfoot}%
\usepackage{booktabs}%
\usepackage{algorithm}%
\usepackage{algorithmicx}%
\usepackage{algpseudocode}%
\usepackage{listings}%
\usepackage{mathdots}

\usepackage[latin1]{inputenc}
\usepackage{indentfirst}
\usepackage{amssymb}
\usepackage{amsfonts}
\usepackage{graphicx}
\usepackage{array}
\usepackage{amsthm}
\usepackage[normalem]{ulem}
\usepackage{indentfirst}
\usepackage{amsmath}               
\usepackage{amsfonts}              
\usepackage{amsthm}                
\usepackage{graphics}
\usepackage{verbatim}
\usepackage{tikz}
\usepackage{hyperref}

\tikzstyle{vertex}=[circle, draw, inner sep=0pt, minimum size=4pt]

\usepackage{xcolor,soul}

\usetikzlibrary{automata, arrows}

\tikzset{main node/.style={circle,fill=blue!20,draw,minimum size=1cm,inner sep=0pt},
            }

\usepackage{mathtools}

\def\Tr{\mathrm{Tr}}
\def\bma{\begin{bmatrix}}
\def\ema{\end{bmatrix}}
\def\bex{\begin{example}}
\def\eex{\end{example}}
\def\beq{\begin{equation}}
\def\eeq{\end{equation}}

\def\qed{\begin{flushright} $\square$ \end{flushright}}
\def\qee{\begin{flushright} $\Diamond$ \end{flushright}}

\DeclarePairedDelimiter\bra{\langle}{\rvert}
\DeclarePairedDelimiter\ket{\lvert}{\rangle}

\newtheorem{teo}{Theorem}[section]
\newtheorem{lema}[teo]{Lemma}
\newtheorem{corollary}[teo]{Corollary}
\newtheorem{defi}[teo]{Definition}
\newtheorem{prop}[teo]{Proposition}

\theoremstyle{definition}

\newtheorem{ex}[teo]{Example}
\newtheorem{REM}[teo]{Remark}

\begin{document}

\title[Article Title]{Recurrence Criteria for Reducible Homogeneous Open Quantum Walks on
the Line and on the Grid}

\author*[1]{\fnm{Newton} \sur{Loebens}}\email{newtonloebens@gmail.com}

\affil*[1]{\orgname{Universidade Federal do Rio Grande do Sul}, \orgaddress{\street{Av. Bento Gon\c{c}alves, 9090 }, \city{Porto Alegre}, \postcode{91540000}, \state{Rio Grande do Sul}, \country{Brazil}}}

\abstract{In this paper, we study the recurrence of Open Quantum Walks induced by finite-dimensional coins on the line ($\mathbb{Z}$) and on the grid ($\mathbb{Z}^2$). Two versions are considered: discrete-time open quantum walks (OQW) and continuous-time open quantum walks (CTOQW). We present three distinct recurrence criteria for OQWs on $\mathbb{Z}$, each adapted to different types of coins. The first criterion applies to coins whose auxiliary map has a unique invariant state, resulting in the first recurrence criterion for Lazy OQWs. The second applies to Lazy OQWs of dimension 2, where we provide a complete characterization of the recurrence for this low-dimensional case. The third one presents a general criterion for finite-dimensional coins in the non-lazy case, which generalizes several of the previously known results for OQWs on $\mathbb{Z}$. Also, we present a general recurrence criterion for OQWs on $\mathbb{Z}^2$ via the open quantum jump chain, obtained from a recurrence criterion for CTOQWs on $\mathbb{Z}^2$.}

\keywords{Homogeneous Open Quantum Walks, Recurrence, Lazy Walks, Coin, Reducibility}

\maketitle

\tableofcontents

\section{Introduction}\label{sec1}


Open Quantum Walks (OQWs) naturally extend classical Markov chains within the framework of open quantum systems. Originally proposed by \cite{attal}, OQWs describe open quantum systems in which repeated measurements of the particle's position, together with the particle's state space, give rise to a Markov chain.

{\color{black}
OQWs have proved to be a versatile framework for describing discrete-time quantum dynamics driven by environmental interaction. Beyond their structural simplicity, they provide an effective language for dissipative implementations of logical operations and state engineering. In particular, it has been shown that single-qubit gates and the CNOT gate can be realized as OQWs on fully connected graphs, and that arbitrary single-qubit states as well as Bell states can be prepared through dissipative mechanisms \cite{attal_graphs}. From a probabilistic viewpoint, central limit theorems (CLTs) have been established for microscopically derived Lazy OQWs \cite{CLTKemp}, as well as for the CTOQWs \cite{bri}.

At the same time, OQWs admit a rich potential-theoretic and asymptotic analysis. Fundamental quantities such as passage times, expected number of visits, recurrence and transience have been investigated \cite{bardet2017passage}. Concerning the implementation of dissipative quantum computing protocols based on the OQW formalism, the approach \cite{effi}  enables faster convergence to the desired steady state and increases the probability of detecting the computational outcome compared to the canonical approach.

More recently a quantum PageRank algorithm based on discrete-time OQWs with Weyl operators as Kraus operators was introduced, demonstrating faster convergence properties than several existing proposals and extending ranking methods to complex networks \cite{dutta}. Recurrence properties of CTOQWs were investigated in \cite{quantum.jump}, where connections with discrete-time OQWs were established and recurrence criteria were derived. Complementarily, \cite{loe_art} analyzed CTOQWs in one dimension through a matrix framework, characterizing site recurrence and related statistics via matrix-valued orthogonal polynomials and Lindblad generators modeling quantum birth-death processes. Finally, \cite{moori} examined hitting-time statistics under quantum channels, providing practical tools to compute first-visit expectations and variances through monitored quantum evolutions and induced absorbing quantum Markov chains.}

In this work, we first consider homogeneous Open Quantum Walks (HOQWs) defined on the set of integers, that is, the graph of this type of walk has the set of sites $V=\mathbb{Z}$, where three linear operators $L, B, R$ correspond to transitions between sites in the quantum walk, remaining constant, with each step moving only to the nearest neighbors $(R, L)$ or staying at the same site $(B)$.

Recurrence is an important probabilistic property of any walk, often characterized by the expected number of returns to the initial site. This is a central concern in the quantum context, and can be directly related to classical recurrence criteria for walks on $\mathbb{Z}$. For example, consider the classical homogeneous walk on $\mathbb{Z}$ with graph as shown in Figure \ref{ClassicGraph}.

\begin{figure}[h!]
$$\begin{tikzpicture}[>=stealth',shorten >=1pt,auto,node distance=2cm]
\node[state] (q-2)      {$-2$};
\node[state]         (q-1) [right of=q-2]  {$-1$};
\node[state]         (q0) [right of=q-1]  {$0$};
\node[state]         (q1) [right of=q0]  {$1$};
\node[state]         (q2) [right of=q1]  {$2$};
\node         (qr) [right of=q2]  {$\ldots$};
\node         (ql) [left of=q-2]  {$\ldots$};
\path[->]          (ql)  edge         [bend left=15]  node[auto] {$r$}     (q-2);
\path[->]          (q-2)  edge         [bend left=15]   node[auto] {$r$}     (q-1);
\path[->]          (q-1)  edge         [bend left=15]   node[auto] {$r$}     (q0);
\path[->]          (q0)  edge         [bend left=15]   node[auto] {$r$}     (q1);
\path[->]          (q1)  edge         [bend left=15]   node[auto] {$r$}     (q2);
\path[->]          (q2)  edge         [bend left=15]   node[auto] {$r$}     (qr);
\path[->]          (qr)  edge         [bend left=15] node[auto] {$l$}       (q2);
\path[->]          (q2)  edge         [bend left=15]   node[auto] {$l$}     (q1);
\path[->]          (q1)  edge         [bend left=15]  node[auto] {$l$}      (q0);
\path[->]          (q0)  edge         [bend left=15]   node[auto] {$l$}     (q-1);
\path[->]          (q-1)  edge         [bend left=15]  node[auto] {$l$}      (q-2);
\path[->]          (q-2)  edge         [bend left=15]  node[auto] {$l$}      (ql);
  \draw [->] (q1) to[loop above]node[auto] {$b$}  (q1);
   \draw [->] (q2) to[loop above]node[auto] {$b$}  (q2);
    \draw [->] (q-2) to[loop above] node[auto] {$b$} (q-2);
     \draw [->] (q-1) to[loop above]node[auto] {$b$}  (q-1);
      \draw [->] (q0) to[loop above]node[auto] {$b$}  (q0);
\end{tikzpicture}$$
\caption{Classical Homogeneous Lazy Random Walk on $\mathbb{Z}$.}
\label{ClassicGraph}
\end{figure}

It is well known that this lazy random walk with $r,l\geq 0,\;b>0,\;l+b+r=1,$ is recurrent if and only if the transition parameters $r$ (rate of jumping to the right) and $l$ (rate of jumping to the left) are equal. Also, if $b=0,$ then the walk is non-lazy and it is recurrent if and only if $r = 1/2 = l$ \cite{norris}.

In the context of discrete-time Markov chains, a state is called recurrent if, once the chain enters that state, the probability of returning to it is 1. This is equivalent to saying that the average number of returns to that state is infinite, indicating that the chain will revisit the state indefinitely \cite{norris}.  {\color{black} The definition of recurrence employed on this work follows the terminology of \cite{bardet2017passage}.
As shown therein, this notion is better suited to the quantum setting than the classical ``probability of return equals 1" criterion, since for OQWs the two are not equivalent. Moreover, for irreducible semifinite OQWs, the average number of returns satisfies a universality property: it is either finite for all vertices and all initial internal states, or infinite for all vertices and all initial internal states. The same does not hold for the probability of return, which may equal 1 for some internal states and be strictly less than 1 for others, even when the walk starts at the same site. The mean return definition also connects naturally to the limit theorems used throughout this work and allows recurrence to be characterized in terms of a drift parameter $m$.}

With this description, it is natural to seek criteria to determine under what conditions an HOQW is recurrent. This problem has been addressed in the literature with the development of specific criteria, which are essential to understanding the dynamics and long-term behavior of these systems. See, for instance,
\cite{CGL,class,LS.gambler}. In this context, recurrence criteria for higher-order HOQWs have been established in \cite{thomas}, where the authors derive such criteria for non-lazy HOQWs with two-dimensional coins as well as for arbitrary finite dimensions in the case of irreducible coins. 

In this article, we present more general criteria for the recurrence of lazy and non-Lazy HOQWs, based on the transition operators $L,B,R$ when acting on finite-dimensional vector spaces, that is, we allow the HOQW to be reducible. Furthermore, we present the first recurrence criterion for Lazy HOQWs, using results obtained from recent studies on CLTs for this kind of HOQWs. A first CLT for OQWs was established in \cite{attal2015central}. In this work, we rely on more recent developments, including the CLT for Lazy HOQWs presented in \cite{CLTKemp}, as well as the general CLT for HOQWs on lattices with finite internal degrees of freedom given in \cite{carboneCLT}.

By making use of the ideas above and generalizing results from the classical version to recurrence, we also obtain a recurrence criterion for the HOQW on the grid, which is the case where the set of vertices is $\mathbb{Z}^2$; however, this is achieved through continuous-time OQWs (CTOQWs). For this purpose, we use recent results of \cite{quantum.jump}, where the introduction of the open quantum jump chain allowed for the association between recurrence in discrete-time and continuous-time versions of OQWs.

{\color{black}This work is structured as follows. In Section 2, we define the OQWs induced by a coin and their transition probabilities. Section 3 is devoted to presenting a recurrence criterion for 
Lazy HOQWs under ergodicity. Section 4 discusses recurrence under ergodicity for discrete-time and continuous-time OQWs on the grid. General recurrence criteria for HOQWs on the line and on the grid are detailed on Section 5. Section 6 illustrates a pack of comprehensive examples.
}

{\color{black}
While a unified criterion encompassing both laziness and reducibility is theoretically desirable, the primary goal of this paper is to elucidate the recurrence properties of lazy walks in the ergodic regime. The effect of reducibility is addressed exclusively in the non-lazy regime. A combined treatment would require a more involved technical framework that could obscure these distinct contributions. For this reason, we have chosen to present them separately, leaving the general lazy reducible case as a natural direction for future research.

For the reader's convenience, we summarize in Table~\ref{tab:discrete} the main recurrence criteria for discrete-time HOQWs established throughout this paper, indicating for each result the graph ($\mathbb{Z}$ or $\mathbb{Z}^2$), the dimension of the internal space, whether the walk is lazy or non-lazy, and whether it applies to the ergodic or general (reducible) case. For continuous-time HOQWs on the grid, the main result is Theorem \ref{ct.criterion}, which provides a recurrence criterion under the assumption that the auxiliary map has at most one minimal enclosure.

\begin{table}[h!]
\centering
\begin{tabular}{|c|c|c|c|c|}
\hline
\textbf{Criterion} & \textbf{Graph} & \textbf{Dimension} & \textbf{Lazy/Non-Lazy} & \textbf{Type} \\
\hline
Theorem \ref{teo_lazy} & $\mathbb{Z}$ & Finite $d$ & Lazy & Ergodic \\
\hline
Theorem \ref{LazyDim2} & $\mathbb{Z}$ & $d = 2$ & Lazy & General \\
\hline
Theorem \ref{TeoGeneralCriteria} & $\mathbb{Z}$ & Finite $d$ & Non-Lazy & General \\
\hline
Theorem \ref{OQWcriterion} & $\mathbb{Z}^2$ & Finite $d$ & Non-Lazy & Ergodic \\
\hline
Theorem \ref{TeoGeneralCriteria2} & $\mathbb{Z}^2$ & Finite $d$ & Non-Lazy & General \\
\hline
\end{tabular}
\caption{Summary of discrete-time recurrence criteria.}
\label{tab:discrete}
\end{table}
}

\section{General Settings}

Let us consider the set of vertices $\mathbb{Z},$ and $\mathcal{K} = \mathbb{C}^\mathbb{Z}$ be the state space of a quantum system with as many degrees of freedom as the number of vertices.  To emphasize the position of the particle of our quantum walk, we consider a quantum system described by the separable complex Hilbert space
\begin{equation*}\label{Hdef}
\mathcal{H}=\bigoplus_{i\in \mathbb{Z}}\mathfrak{h}_i,
\end{equation*}
where the $\mathfrak{h}_i$ are separable Hilbert spaces. The label $i\in \mathbb{Z}$ represents the position of the particle and, when the
particle is located at the vertex $i\in \mathbb{Z}$, its internal state is encoded in the space $\mathfrak{h}_i,$ as it will be detailed below.

\subsection{Lazy Open Quantum Walks}
We consider the set of density operators over a finite Hilbert space $\mathcal{W},$
$$\mathcal{D}(\mathcal{W})=\{\rho:\mathcal{W}\rightarrow \mathcal{W}:\; \rho\geq 0,\; \mathrm{Tr}(\rho)=1\},$$
and introduce
$$
\mathcal{S}(\mathcal{H}\otimes \mathcal{K}) =
\left\{\tau\in\mathcal{I}(\mathcal{H}\otimes\mathcal{K}) :
\textstyle \tau=\displaystyle\sum_{j\in \mathbb{Z}}\rho(j)\otimes\ket{j}\bra{j},\quad \rho(j)\geq 0,\quad\sum_{j\in\mathbb{Z}}\Tr\left(\rho(j)\right)=1\right\},
$$
the set of density operators acting on $\mathcal{H}\otimes \mathcal{K}.$ {\color{black} Also, given a vector space $T$, we denote the set of bounded operators acting on $T$ by $\mathcal{B}(T)$.}

We are interested in HOQWs of the form
\beq\label{oqwh1}\Phi(\tau)=\sum_{i\in\mathbb{Z}}\Big(L\rho(i+1) L^*+B\rho(i) B^*+R\rho(i-1) R^* \Big)\otimes |i\rangle\langle i|,\;\;\;\tau\in \mathcal{S}(\mathcal{H}\otimes\mathcal{K}),\eeq
where $L,B,R\in\mathbb{M}_d({\mathbb{C}})$ satisfy $L^*L+B^*B+R^*R=I_d.$ 

We say that the triple of matrices $(L,B,R)$ is a \textbf{coin} if eq. \eqref{oqwh1} holds. Therefore, we can assume that $\mathfrak{h}_i=\mathfrak{h}\;\forall i\in \mathbb{Z},$ and if $\mathfrak{h}=\mathbb{C}^d$ for some $d$, the \textbf{dimension} of the coin is said to be $d$. In this article, we consider only the finite-dimensional case $d<\infty.$

If $B\neq 0,$ the quantum walk exhibits a lazy behavior, meaning that in the context of quantum walks, there is a non-zero probability for the walker to stay in the same vertex after a step. We refer to the HOQW above as a \textbf{HOQW induced by the coin $(L,B,R)$} in which the walk is inherently lazy.

Therefore, this coin represents a homogeneous, nearest-neighbor OQW on the integer line with the left and right transitions given by $L$ and $R$, respectively, while $B$ describes the rate of no jump at some instant. This lazy behavior introduces distinct properties compared to its non-lazy counterpart. Notably, the presence of the matrix $B$ influences the walk's propensity to remain in its current vertex, thereby affecting the overall dynamics. If $B=0,$ then $\Phi$ will be called a \textbf{HOQW induced by a coin $(L,R),$} or a \textbf{non-Lazy HOQW induced by a coin $(L,R).$} The dynamic described above is represented by the graph in Figure \ref{originalOQWonZ}.

\begin{figure}[h!]
$$\begin{tikzpicture}[>=stealth',shorten >=1pt,auto,node distance=2cm]
\node[state] (q-2)      {$-2$};
\node[state]         (q-1) [right of=q-2]  {$-1$};
\node[state]         (q0) [right of=q-1]  {$0$};
\node[state]         (q1) [right of=q0]  {$1$};
\node[state]         (q2) [right of=q1]  {$2$};
\node         (qr) [right of=q2]  {$\ldots$};
\node         (ql) [left of=q-2]  {$\ldots$};
\path[->]          (ql)  edge         [bend left=15]  node[auto] {$R$}     (q-2);
\path[->]          (q-2)  edge         [bend left=15]   node[auto] {$R$}     (q-1);
\path[->]          (q-1)  edge         [bend left=15]   node[auto] {$R$}     (q0);
\path[->]          (q0)  edge         [bend left=15]   node[auto] {$R$}     (q1);
\path[->]          (q1)  edge         [bend left=15]   node[auto] {$R$}     (q2);
\path[->]          (q2)  edge         [bend left=15]   node[auto] {$R$}     (qr);
\path[->]          (qr)  edge         [bend left=15] node[auto] {$L$}       (q2);
\path[->]          (q2)  edge         [bend left=15]   node[auto] {$L$}     (q1);
\path[->]          (q1)  edge         [bend left=15]  node[auto] {$L$}      (q0);
\path[->]          (q0)  edge         [bend left=15]   node[auto] {$L$}     (q-1);
\path[->]          (q-1)  edge         [bend left=15]  node[auto] {$L$}      (q-2);
\path[->]          (q-2)  edge         [bend left=15]  node[auto] {$L$}      (ql);
  \draw [->] (q1) to[loop above]node[auto] {$B$}  (q1);
   \draw [->] (q2) to[loop above]node[auto] {$B$}  (q2);
    \draw [->] (q-2) to[loop above] node[auto] {$B$} (q-2);
     \draw [->] (q-1) to[loop above]node[auto] {$B$}  (q-1);
      \draw [->] (q0) to[loop above]node[auto] {$B$}  (q0);
\end{tikzpicture}$$
\caption{Lazy HOQW on $\mathbb{Z}$.}
\label{originalOQWonZ}
\end{figure}

The \textbf{quantum trajectory} of this kind of HOQW on the line, induced by a coin $(L,B,R)$ and starting from a state $\tau$ of the form $\tau=\sum_{i \in \mathbb{Z}}\tau_i \otimes |i\rangle\langle i|,$
is any path generated by the Markov chain $(X_n,\tau_n)_{n\geq 0}$, where $X_n$ denotes the position of the particle at time $n$ and $\tau_n$ its internal degree. The transition probabilities are given by
\begin{equation}\nonumber
\begin{split}
  \mathbb{P}\left(\, (X_{n+1},\tau_{n+1}) = \left(i-1, \frac{L\sigma L^*}{\mathrm{Tr}(L\sigma L^*)}\right) \; \Bigg| \; (x_n,\tau_n) = (i, \sigma)\,\right) &= \mathrm{Tr}(L\sigma L^*),  \\
  \mathbb{P}\left(\, (X_{n+1},\tau_{n+1}) = \left(i, \frac{B\sigma B^*}{\mathrm{Tr}(B\sigma B^*)}\right) \; \Bigg| \; (x_n,\tau_n) = (i, \sigma)\,\right) &= \mathrm{Tr}(B\sigma B^*),  \\
  \mathbb{P}\left(\, ( X_{n+1},\tau_{n+1}) = \left(i+1, \frac{R\sigma R^*}{\mathrm{Tr}(R\sigma R^*)}\right) \; \Bigg| \; (x_n,\tau_n) = (i, \sigma)\,\right)& = \mathrm{Tr}(R\sigma R^*),
\end{split}
\end{equation}
for every $i \in \mathbb{Z}$, $\sigma \in \mathcal{D}(\mathfrak{h})$, and initial law
$$
\mathbb{P}\left(\, (X_{0},\tau_0) = \left(i, \frac{\tau_i}{\mathrm{Tr}\,\tau_i}\right) \; \right) = \mathrm{Tr}(\tau_i),
$$
and all other transition probabilities are null. Therefore, at time \( n \), if the system is located at site \( \ket{i} \) with an initial density \( \rho\otimes\ket{i}\bra{i} \), then at the next time step \( n+1 \), one of the following occurs:  

\begin{itemize}
\item The particle moves to site \( \ket{i-1} \) (left) with probability \( \operatorname{Tr}(L_i \rho L_i^*) \), and the density evolves to  
  \[
  \frac{L_i \rho L_i^*}{\operatorname{Tr}(L_i \rho L_i^*)} \otimes \ket{i-1} \bra{i-1};
  \]

\item The particle remains at site \( \ket{i} \) with probability \( \operatorname{Tr}(B_i\rho B_i^*) \), and the new density matrix is  
  \[
  \frac{B_i \rho B_i^*}{\operatorname{Tr}(B_i \rho B_i^*)} \otimes \ket{i} \bra{i};
  \]

\item The particle moves to site \( \ket{i+1} \) (right) with probability \( \operatorname{Tr}(R_i\rho R_i^*) \), leading to the new density matrix  
  \[
  \frac{R_i \rho R_i^*}{\operatorname{Tr}(R_i \rho R_i^*)} \otimes \ket{i+1} \bra{i+1}.
  \]
\end{itemize}

{\color{black}

While irreducibility has been the focus of previous studies on the asymptotic behavior of HOQWs, this work extends the analysis to the reducible case. We now recall the distinction between these two notions. 

We consider a coin $(L,B,R)$ and its \textbf{auxiliary map} $\mathcal{L}:\mathcal{B}(\mathfrak{h})\rightarrow \mathcal{B}(\mathfrak{h})$ given by
$$
\mathcal{L}(\rho)=L\rho L^*+
B\rho B^*+R\rho R^*,\quad \rho\in \mathcal{B}(\mathfrak{h}).
$$
Note that $\mathcal{L}$ is a quantum channel, thus it has at least one invariant state, that is, there exists some density operator $\rho_\infty\in \mathcal{B}(\mathfrak{h})$ satisfying $\mathcal{L}(\rho_\infty)=\rho_\infty.$ When the invariant state of $\mathcal{L}$ is unique, we will say that the map $\mathcal{L}$ is \textbf{ergodic}.

As in classical Markov chains, an irreducible HOQW cannot be decomposed into dynamically independent subsystems, ensuring well-defined ergodic properties \cite{HOM-CaPautrat}. 

\begin{prop}\label{4.1CP}\cite[Proposition 4.1]{HOM-CaPautrat}
Let $\Phi$ be an HOQW defined by transition operators $L_s$, 
and $\mathcal{L}$ its auxiliary map.
\begin{enumerate}
\item $\mathcal{L}$ is irreducible if and only if the operators $\{L_s, s\in S\}$ have no nontrivial closed invariant subspace in common.
\item $\Phi$ is irreducible if and only if all possible concatenations of transitions starting and ending at the same site have no nontrivial closed invariant subspace in common.
\end{enumerate}
\end{prop}

A direct consequence is the following:

\begin{corollary}\cite[Corollary 4.2]{HOM-CaPautrat}
If $\Phi$ is irreducible, then $\mathcal{L}$ is irreducible.
\end{corollary}

This corollary implies that if the auxiliary map $\mathcal{L}$ is reducible, then the HOQW $\Phi$ is also reducible. As noted in~\cite[Remark 3.6]{HOM-CaPautrat}, an irreducible map admits at most one invariant state, and if such a state exists it must be faithful. Conversely, if a map has a unique faithful invariant state, then it is irreducible. This connection between algebraic structure and spectral properties provides a practical way to verify irreducibility in concrete examples.

For instance, let  $\mathfrak{h}=\mathbb{C}^2$, $S=\{-1,+1\}$, and denote $L_{-1}=L$, $L_{+1}=R,$ where 
\[
L = \frac{1}{\sqrt{20}}\begin{bmatrix} 4 & -1 \\ 0 & 2 \end{bmatrix}, \qquad
R = \frac{1}{\sqrt{20}} \begin{bmatrix} 2 & 2 \\ 0 & \sqrt{11} \end{bmatrix}.
\]
The auxiliary map $\mathcal{L}$ has a unique invariant state $\rho_\infty=\begin{bmatrix}
    1&0\\0&0
\end{bmatrix},$
which is not faithful. Therefore, $\mathcal{L}$ is reducible, and consequently $\Phi$ is also reducible.

}
\section{Recurrence Criteria for Lazy Open Quantum Walks}

\subsection{Faithfulness of Density Operators}
Density operators play a fundamental role in open quantum systems. In this work, they provide statistical descriptions of HOQWs. Faithful and non-faithful density operators are characterized by their positive definiteness and positive semi-definiteness, respectively, when they evolve in finite-dimensional systems.

Faithful density operators are defined as being positive definite. This means that the eigenvalues of a faithful density operator are strictly positive, and the operator is invertible. Essentially, faithfulness ensures a complete representation of the quantum system, capturing the full richness of its physical properties.

On the other hand, non-faithful density operators are positive semi-definite, thus they have $\lambda = 0$ as one of their eigenvalues. The presence of zero eigenvalues implies that the non-faithful density operator is not invertible and may represent specific mixed states or situations where information loss occurs. The positive semi-definiteness criterion relaxes the strictness of positive definiteness, allowing for a broader range of quantum state representations, while still maintaining the one-to-one correspondence between quantum states and density operators.

An important subset of non-faithful density operators is composed of the pure density operators. A \textbf{pure density} is a density $\rho$ that is a projection of rank 1, i.e., there exists some vector $\ket{u}\in\mathbb{C}^d$ such that $\rho =\ket{u}\bra{u}$.

Let us consider an HOQW starting the walk at vertex $\ket{i}$ with initial density operator $\rho.$ For evolution at time $n=0,1,2,\ldots,$ we define
\begin{equation*}
p_n(j)=\mathbb{P}(\mbox{the quantum particle, at time }n,\mbox{ is in site}\;\ket{j})=\mathrm{Tr}(\rho_{n}(j)),
\end{equation*}
where
$$
\Phi^n(\rho\otimes\ket{i}\bra{i})=\sum_{s\in \mathbb{Z}}\rho_n(s)\otimes\ket{s}\bra{s}.
$$

With this notation,
\begin{equation*}\label{ctoqwprobmat}
p_{ji;\rho}(n)=p_{n}(\rho\otimes |i\rangle \to |j\rangle)=\mathrm{\mathrm{Tr}}(\rho_n(j)\otimes\ket{j}\bra{j})=\mathrm{\mathrm{Tr}}\left(\Phi^n(\rho\otimes\ket{i}\bra{i})(I\otimes\ket{j}\bra{j})\right)
\end{equation*}
denote the probability of being at site $\ket{j}$ at step $n,$ given that we started at site $\ket{i},$ with initial density $\rho$ concentrated at $\ket{i}.$  Therefore, the dynamics starts with a density operator $\rho$ concentrated at some vertex $\ket{i},$ takes the evolution up to time ($n$) through the linear operator $\Phi^n,$ producing a new density operator
$$\rho_n=\sum_s\rho_n(s)\otimes\ket{s}\bra{s}=\Phi^n(\rho\otimes\ket{i}\bra{i}),\quad \mathrm{Tr}\left(\sum_s\rho_n(s)\right)=1,$$
and then we project $\rho_n$ onto the subspace generated by vertex $\ket{j}.$ Note that
$$
\Phi^n(\rho\otimes\ket{i}\bra{i})(I\otimes\ket{j}\bra{j})=\rho_n(j)\otimes\ket{j}\bra{j},
$$
which represents the data concentrated at vertex $\ket{j}$ at time $n.$

Consider an HOQW on some set of vertices $\mathbb{Z}$, let $i\in \mathbb{Z}$ and $\rho\in \mathcal{D}(\mathfrak{h})$. We say that a vertex $\ket{i}$ is
\begin{enumerate}
  \item $\rho$-\textbf{recurrent}\index{$\rho$-recurrent vertex} if
  \begin{equation}\label{prec}
    {\color{black}\mathbb{E}_{i,\rho}(n_i):=}\sum_{n=0}^{\infty}p_{ii;\rho}(n)=\infty,\quad{\color{black}\text{ where } p_{ii;\rho}(0)=1.}
  \end{equation}
Otherwise, $\ket{i}$ is said to be $\rho$-\textbf{transient};\index{$\rho$-transient vertex}
    \item \textbf{recurrent},\index{recurrent vertex} if $\ket{i}$ is $\rho-$recurrent for all $\rho\in \mathcal{D}(\mathfrak{h});$
  \item \textbf{transient},\index{transient vertex} if $\ket{i}$ is $\rho-$transient for all $\rho\in \mathcal{D}(\mathfrak{h}).$
  \end{enumerate}

Concerning all vertices, an OQW is called: \textbf{recurrent} if every vertex is recurrent; \textbf{transient} if every vertex is transient. Note that equation \eqref{prec} computes the mean number of returns to site $\ket{i}.$ {\color{black}
The choice of this definition, based on the mean number of returns, rather than the probability of return equals $1$ ($\mathbb{P}_{i,\rho}(t_i < \infty) < 1$, where $t_i$ is the returning time to $i$) definition, requires a careful justification. As demonstrated in \cite{bardet2017passage}, for OQWs these two notions are not equivalent, unlike the classical case. This non-equivalence makes a conceptual choice necessary regarding which definition to adopt.

The definition adopted in this work, based on the mean number of returns $\mathbb{E}_{i,\rho}(n_i)$, follows the terminology proposed in \cite{bardet2017passage} for semifinite irreducible OQWs. As established in \cite[Corollary 3.10]{bardet2017passage}, for irreducible semifinite OQWs the value of $\mathbb{E}_{i,\rho}(n_i)$ is universal: it is either finite for every $i \in V$ and every $\rho \in \mathcal{D}(\mathfrak{h}_i)$, or infinite for every $i$ and every $\rho$. This universality does not hold for the return probability $\mathbb{P}_{i,\rho}(t_i < \infty)$, which may assume different values for different internal states even in irreducible OQWs (as illustrated in  \cite[Example 5.2]{bardet2017passage}). Therefore, the definition based on the mean number of visits is more suitable for a robust and consistent classification of asymptotic behavior, connecting naturally to the limit theorems used throughout this work and allowing the characterization of recurrence via the drift parameter $m$, as developed in the following sections.

It is worth recalling that in the particular case of irreducible HOQWs, Jacq and Lardizabal \cite{thomas} demonstrated that these two notions of recurrence do coincide. However, their proof relies heavily on irreducibility assumptions, particularly on the uniqueness and faithfulness of the invariant state, which forces the equivalence between the divergence of the mean number of returns and the certainty of return. For reducible HOQWs, the relationship between these two definitions remains an open problem. The existence of multiple invariant states and the possibility of nontrivial transient components within the internal space $\mathfrak{h}$ may lead to situations where a vertex is $\rho$-recurrent under one definition but not the other, depending on the initial internal state. This question, while beyond the scope of the present work, represents an interesting direction for future investigation.
}

Having clarified the choice of recurrence definition and its nuances, we now proceed to the following proposition, which describes $\rho$-recurrence in a concentrated vertex $i\in \mathbb{Z}$ when assumed that $\ket{i}$ is recurrent for some density and $\mathfrak{h}$ has a finite internal degree of freedom. {\color{black} This result establishes the behavior of $\rho$-recurrence when recurrence occurs for at least one initial state. In the following sections, we will build upon this to analyze the structure of the internal space $\mathfrak{h}$ and understand how it decomposes in order to develop recurrence criteria for reducible OQWs.} An immediate consequence will also be given for the case where $\ket{i}$ is $\sigma$-transient for some non-faithful density. 

\begin{teo}\label{discussrec}
Consider an OQW on some set of vertices $V,$ let $i\in V$ and assume that $\mathfrak{h}$ is finite-dimensional.  If $\ket{i}$ is $\tau$-recurrent for some $\tau\in\mathcal{D}(\mathfrak{h}),$ then
\begin{enumerate}
  \item  vertex $\ket{i}$ is $\rho$-recurrent for all faithful $\rho\in\mathcal{D}(\mathfrak{h})$;
\item there exists a pure density operator $\ket{w}\bra{w}$ such that the OQW is $\ket{w}\bra{w}$-recurrent, where $\ket{w}$ is an unit vector of $\mathfrak{h}.$
\end{enumerate}
\begin{proof}
Let $\rho\in \mathcal{D}(\mathfrak{h})$ be faithful. Since the state space \( \mathcal{D}(\mathfrak{h}) \) is compact in the trace-norm topology, the function  
\[
f(\tau) = \inf \{ c > 0 \mid \rho > c\tau \}
\]  
is upper semicontinuous and thus attains its minimum and maximum. In particular, there exists \( c > 0 \) such that  
\[
\rho > c\tau, \quad \forall \rho \in \mathcal{D}(\mathfrak{h}).
\]
Thus
\begin{equation*}
\begin{split}
\sum_{n=0}^{\infty}p_{ii;\rho}(n)=&\sum_{n=0}^{\infty}\textmd{Tr}\left[\Phi^{n}\left(\rho\otimes\ket{i}\bra{i}\right)\right(I\otimes\ket{i}\bra{i})]\\
>&
c\sum_{n=0}^{\infty}\textmd{Tr}\left[\Phi^{n}\left(\tau\otimes\ket{i}\bra{i}\right)(I\otimes\ket{i}\bra{i})\right]\\
=&c\sum_{n=0}^{\infty}p_{ii;\tau}(n)\\
=&\infty,
\end{split}
\end{equation*}
giving the first item.

\medskip

For the second one, let 
$$\tau=\sum_{j}\lambda_j\ket{e_j}\bra{e_j},$$
where each $\lambda_j> 0$ is an eigenvalue of $\tau$ with corresponding eigenvector $\ket{e_j}.$ Then
$$\infty=
\sum_{n=0}^{\infty}\textmd{Tr}\left[\Phi^{n}\left(\tau\otimes\ket{i}\bra{i}\right)\right(I\otimes\ket{i}\bra{i})]=\sum_{j}\left(\lambda_j\sum_{n=0}^{\infty}
\textmd{Tr}\left[\Phi^{n}\left(\ket{e_j}\bra{e_j}\otimes\ket{i}\bra{i}\right)\right(I\otimes\ket{i}\bra{i})]\right),
$$
where at least one term in the sum must be infinite. This completes the proof.
\end{proof}
\end{teo}

By a contraposition argument, we obtain the following direct consequence of Theorem \ref{discussrec}.

\begin{corollary}\label{cor2}
Let $\Phi$ be an OQW on some set of vertices $V,$ $i\in V,$ and $\dim(\mathfrak{h})<\infty.$ If $\ket{i}$ is $\tau$-transient for some faithful  $\tau\in\mathcal{D}(\mathfrak{h}),$ then $\ket{i}$ is transient.
\end{corollary}

Another direct consequence of Theorem \ref{discussrec} is related to the homogeneous case.

\begin{corollary}
Let $\Phi$ be an HOQW on $\mathbb{Z}$ with $\dim(\mathfrak{h})=d<\infty.$ If $\Phi$ is $\tau$-recurrent for some $\tau\in\mathcal{D}(\mathfrak{h}),$ then
\begin{enumerate}
  \item  $\Phi$ is $\rho$-recurrent for all faithful $\rho\in\mathcal{D}(\mathfrak{h})$;
  \item there exists a pure density operator $\ket{w}\bra{w}$ such that $\Phi$ is $\ket{w}\bra{w}$-recurrent, where $\ket{w}$ is an unit vector of $\mathbb{C}^d.$
\end{enumerate}
\end{corollary}

\subsection{Enclosures and Invariant States}

A subspace $\mathcal{Y}$ of $\mathfrak{h}$ is called an \textbf{enclosure} for a quantum channel $\mathcal{L}$ if for every  $\sigma\in \mathcal{D}(\mathfrak{h}),$ we have
$$\mathrm{supp}(\sigma)\subset \mathcal{Y}\Rightarrow \mathrm{supp}\left(\mathcal{L}(\sigma)\right)\subset \mathcal{Y}.$$
If an enclosure contains no other non-trivial enclosures, then it is called a \textbf{minimal enclosure}.

A minimal enclosure with finite dimension is the support of a unique invariant state of the channel, that is, a density operator $\sigma$ such that $\mathcal{L}(\sigma)=\sigma$ {\color{black}\cite[Section
2.2]{carbone-absorption}}.  For more on enclosures of HOQWs, we refer the reader to \cite{carbone-absorption,carboneCLT,bardet2017passage,girotti-thesis}.

Following this property and using the homogeneity of HOQWs induced by a coin, we split $\mathfrak{h}$ into two subsets $\mathcal{Y}$ and $\mathcal{X}$ in such a way that the walk is transient for $\mathcal{D}(\mathcal{Y})$ and recurrent everywhere else, with $\mathcal{Y}$ being an enclosure of the quantum channel with Kraus operators $L,B,R.$

\begin{REM}
From now on we disregard the trivial cases where at least one of the operators $L,B,R$ has some eigenvalue $\lambda$ of modulus 1. To clarify, let $T\ket{v}=\lambda\ket{v},$ where $\ket{v}$ is an unitary vector and $T\in\{L,B,R\}.$ In this case, starting the walk with density operator $\rho=\ket{v}\bra{v},$ the walker is either absorbed by a vertex $(T=B)$ or the particle always jumps to the same side $(T=L$ or $T=R).$
\end{REM}

\begin{prop}\label{splitprop}
Consider an HOQW induced by a coin $(L,B,R)$ of dimension $d$. We are in one (and only one) of the following situations:
\begin{itemize}
  \item $(L,B,R)$ is recurrent;
  \item $(L,B,R)$ is transient;
  \item $(L,B,R)$ is recurrent with respect to all densities except for a convex set $\mathcal{D}(\mathcal{Y}).$ In this case, $\mathcal{Y}$ is an enclosure of $\mathcal{L},$
\begin{equation}\label{C=Y+X}
\mathfrak{h}=\mathcal{Y}\oplus \mathcal{X}
\end{equation}
for some subspace $\mathcal{X}$ of $\mathfrak{h},$ with $\mathcal{Y},\mathcal{X}\neq \{0\}.$
\end{itemize}
\begin{proof} We claim $(L,B,R)$ is $\rho$-transient if and only if it is $\rho_n$-transient for every $n,$ where $\rho_n$ is the density at the $n$-th step of the walk. 

Indeed,
let us suppose the walk starts at vertex $\ket{0}$ with initial density $\rho.$ For instance, if the first jump is to $\ket{-1},$ the density, after this first jump, is
$$
\rho_1=\frac{L\rho L^*}{\Tr(L\rho L^*)}.
$$
Since the walk is homogeneous, it is $\rho$-transient if and only if it is $\rho_1$-transient. After $n$ jumps, the density is
\begin{equation}\label{Jpi}
\rho_n=\frac{J_\pi\rho J_\pi^*}{\Tr(J_\pi\rho J_\pi^*)},
\end{equation}
where $J_\pi$ is a concatenation of $n$ operators, each one belonging to $\{L,B,R\},$ proving the claim, once this is valid for any $n=1,2,3,\ldots$

Moreover, by equation \eqref{Jpi} we have that {\color{black}span$\left(\cup_n\;\text{supp}(\rho_n)\right)\subset\mathfrak{h}.\;$} Joining all the possible initial states in which $(L,B,R)$ is transient, we obtain a decomposition \eqref{C=Y+X}, where $\mathcal{Y}$ or $\mathcal{X}$ may be $\{0\}$, and by definition $\mathcal{Y}$ is an enclosure.

Let us suppose that $(L,B,R)$ is neither recurrent nor transient. By Theorem \ref{discussrec}, there exist unit vectors $\ket{u},\ket{v}\in\mathfrak{h}$ such that the walk is $\ket{u}\bra{u}$-recurrent and $\ket{v}\bra{v}$-transient. Let $\mathcal{Y}$ be the subspace of $\mathfrak{h}$ such that $\mathcal{D}(\mathcal{Y})$ is the set of densities of $\mathcal{B}(\mathfrak{h})$ in which $(L,B,R)$ is transient, thus we pick $\rho\in \mathcal{D}(\mathcal{Y})$ and we can affirm that $(L,B,R)$ is $\rho_n$-transient for every $n.$ Finally, take $\mathcal{X}=\mathcal{Y}^\perp$ and the trichotomy holds.
\end{proof}
\end{prop}

\subsection{Recurrence under Ergodicity}

{\color{black}
Following the CLT approach for lazy OQWs given in \cite{CLTKemp}, we consider the Markov chain
$(\rho_n,\Delta X_n)_{n\geq 0}$ associated with the quantum trajectory description of the walk. Let $m\in\mathbb{R}^d$ denote the
drift vector of the walk, given by the expected value of the position increment under the invariant
state $\rho_\infty$, namely
\[
m=\sum_{k} \mathrm{Tr}(A_k \rho_\infty A_k^*)\, e_k ,
\]
where $(e_k)_k$ are the possible displacement vectors of the walk, $A_k$ are the transition operators. For a fixed vector $l\in\mathbb{R}^d$, the quantity $(X_n-nm)\cdot l$ admits a Doob decomposition
into the sum of a martingale part and a predictable process. The martingale term is given by
\begin{equation}\label{eq27}
M_n=\sum_{j=2}^{n}\left[f(\rho_j,\Delta X_j)-Pf(\rho_{j-1},\Delta X_{j-1})\right],
\end{equation}
where $f:\mathcal{D}(\mathfrak{h})\times \mathbb{R}^d \to \mathbb{R}$ is defined by
\[
f(\rho,x)=\mathrm{Tr}(\rho L_l)+x\cdot l,
\]
and $L_l=L\cdot l$ denotes the operator associated with the vector $l\in\mathbb{R}^d$ acting on the internal Hilbert space $\mathfrak{h}$.
Here $\rho_j$ denotes the internal state of the walker at time $j$, $\Delta X_j=X_j-X_{j-1}$ represents
the increment of the position process, and $P$ is the transition operator of the Markov chain
$(\rho_n,\Delta X_n)$.
}

\begin{teo}\label{LawOfLargeNumbers}
Consider a Lazy HOQW induced by a coin $(L,B,R)$ of dimension $d.$ If the auxiliary map $\mathcal{L}$ admits only one invariant state $\rho_{\infty},$ then
\begin{equation*}\label{LLNeq}
\frac{X_n}{n}\stackrel{n\rightarrow\infty}{\longrightarrow} m,\qquad \text{where }\qquad m=\mathrm{Tr}\left(R\rho_\infty R^*\right)-\mathrm{Tr}\left(L\rho_\infty L^*\right),
\end{equation*}
in probability.
\end{teo}
\begin{proof}
This result can be obtained by using the Doob decomposition as described in Equation \eqref{eq27} and \cite[Theorem 5.2]{attal2015central} for the Martingale $M_n$ satisfying
$$
X_n-n.m=X_0+M_n+\beta_n,
$$
where $M_n/n\rightarrow 0$ and $\beta_n\rightarrow 0$ when $n\rightarrow {\color{black}\infty}.$

\end{proof}

\begin{teo}[Chung-Fuchs Theorem for HOQWs with ergodic auxiliary map]\label{chungF}
Consider an HOQW induced by a coin $(L,B,R),\;\mathcal{L}$ its auxiliary map and $\mathcal{Y}$ an enclosure. Suppose that the initial density operator is $\rho\in\mathcal{D}(\mathcal{Y})$ with quantum trajectories $(X_n,\rho_n)_{n\geq 0}$ and $X_n/n\rightarrow 0$ in probability, then $(L,B,R)$ is $\sigma$-recurrent for some $\sigma\in \mathcal{D}(\mathcal{Y}).$
\begin{proof}
The proof is analogous to \cite[Theorem 7]{thomas} and we will proceed by remarking on how it can be generalized. Firstly, the maximum appearing in \cite[Lemma 6]{thomas} can be taken in $\mathcal{D}=\mathcal{D}(\mathcal{Y}).$ {\color{black} Indeed, if the initial state is $\rho\in\mathcal{D}(\mathcal{Y}),$ then the density operator $\rho_n$ defined in equation \eqref{Jpi} lies on $\mathcal{D}(\mathcal{Y})$  for every $n,$ otherwise $\mathcal{Y}$ would not be an enclosure, by definition, thereby the maximum taken in the proof of \cite[Lemma 6]{thomas} can be reduced to $\mathcal{D}(\mathcal{Y}).$} Therefore, by the same reason, the density operator considered in \cite[Theorem 7]{thomas} belongs to $\mathcal{D}(\mathcal{Y}),$ proving the Theorem for the non-lazy case. Moreover, the lazy case holds as an extension of this proof, since all the results work for HOQWs on $\mathbb{Z}$.
\end{proof}
\end{teo}

The following theorem presents the first recurrence criterion for non-irreducible HOQWs of arbitrary finite dimension. This result extends \cite[Corollary 8 and Proposition 8]{thomas} for reducible CTOQWs and can also be applied to Lazy HOQWs.

\begin{teo}[Recurrence Criteria]\label{teo_lazy}
Let us consider an HOQW on $\mathbb{Z}$ induced by a coin $(L,B,R)$ of dimension $d$ such that the auxiliary map $\mathcal{L}$ is ergodic (hence $\mathcal{L}$ has a unique invariant state $\rho_{\infty}$). {\color{black} Under this assumption, the mixed case described in Proposition \ref{splitprop} cannot occur, and we have the following dichotomy:}
\begin{itemize}
  \item $\textmd{Tr}\left(L\rho_\infty L^*\right)=\textmd{Tr}\left(R\rho_\infty R^*\right)\Rightarrow$ the walk is recurrent;
  \item $\textmd{Tr}\left(L\rho_\infty L^*\right)\neq\textmd{Tr}\left(R\rho_\infty R^*\right)\Rightarrow$ the walk is transient.
\end{itemize}
\begin{proof}
We suppose that the walk is neither recurrent nor transient. In this case, the walk is recurrent for some density and transient for all densities in some $\mathcal{D}(\mathcal{Y}),$ where $\mathcal{Y}$ is an enclosure (Proposition \ref{splitprop}). Moreover, the unique invariant state $\rho_\infty$ of $\mathcal{L}$ must belong to $\mathcal{D}(\mathcal{Y}).$ 

If $\;\textmd{Tr}\left(L\rho_\infty L^*\right)=\textmd{Tr}\left(R\rho_\infty R^*\right),$ then $X_n/n\rightarrow 0$ by Theorem \ref{LawOfLargeNumbers}, thus $(L,B,R)$ is $\rho$-recurrent for some $\rho\in\mathcal{D}(\mathcal{Y})$ by Theorem \ref{chungF}, which is a contradiction.

On the other hand, if $\textmd{Tr}\left(L\rho_\infty L^*\right)\neq\textmd{Tr}\left(R\rho_\infty R^*\right),$ then $X_n/n\rightarrow m\neq 0.$ Therefore  $\mathbb{P}_{0,\rho}(X_n=0\mbox{ i.o.})=0,$ showing that the mean number of returns to the initial site is finite for any $\rho$, thus the walk is transient,
which is also a contradiction. We have shown that either the walk is recurrent or transient.

Therefore, if $\textmd{Tr}\left(L\rho_\infty L^*\right)=\textmd{Tr}\left(R\rho_\infty R^*\right),$ then the walk is $\rho$-recurrent for some density and thus it is recurrent. If $\textmd{Tr}\left(L\rho_\infty L^*\right)\neq\textmd{Tr}\left(R\rho_\infty R^*\right),$ then the walk is transient for every density. 
\end{proof}
\end{teo}
{\color{black}Note that the mixed case from Proposition \ref{splitprop} does not occur on Theorem \ref{teo_lazy}, as ergodicity ensures a strict dichotomy.}

We remark that the operator $B$ does influence in the recurrence of $(L,B,R).$ Indeed, even it does not appear on the bullet points that describe the recurrence criterion, the invariant state $\rho_\infty$ may change when we pick distinct $B$'s. See example  \ref{ex3} below. {\color{black}In addition to, a distinction must be drawn between lazy and non-lazy walks regarding their recurrence behavior. In the non-lazy case ($B = 0$), the walker is forced to jump at every step, and recurrence is determined solely by the balance between left and right transitions~\cite{thomas}. In the lazy case ($B \neq 0$), the presence of a nonzero ``stay'' probability introduces additional freedom: the walker may remain at the same site, which affects the asymptotic drift and the structure of invariant states. Consequently, recurrence criteria for lazy walks involve the full triple $(L,B,R)$, while non-lazy criteria depend only on $(L,R)$. This distinction is reflected throughout  Theorem~\ref{teo_lazy}, which provides a criterion for lazy walks under ergodicity.

It is worth comparing our results with those of~\cite{thomas}. In that work, the authors established recurrence criteria for irreducible OQWs in the non-lazy setting, with recurrence occurring precisely when $\Tr(L\rho_\infty L^*) = \Tr(R\rho_\infty R^*) = 1/2$. Our Theorem~\ref{teo_lazy} extends this analysis to lazy walks under the assumption that the auxiliary map $\mathcal{L}$ is ergodic. In this context, recurrence is determined by the condition $\Tr(L\rho_\infty L^*) = \Tr(R\rho_\infty R^*)$, without requiring this common value to be $1/2$. The value $1/2$ in~\cite{thomas} is a consequence of the irreducibility assumption and the particular normalization used there. By taking $B = 0$ and assuming irreducibility in our framework, we recover the criterion of~\cite{thomas} as a special case.}

\section{2D Open Quantum Walks}

We start this section with the continuous-time OQW.

\subsection{Continuous-Time Open Quantum Walks}
In order to define the CTOQWs, we recall that
an operator \textbf{semigroup}\index{semigroup} $\mathcal{T}$ on a Hilbert space $\mathcal{K}$ is a family of bounded linear operators $(T_t)$ acting
on $\mathcal{K}$, $t\geq 0,$ such that
\begin{equation}\label{def-semi}\nonumber
T_tT_s=T_{t+s},\;\forall s,t\in\mathbb{R}_+\;\mbox{ and }\;T_0=I_\mathcal{K}.
\end{equation}

If $t\mapsto T_t$ is continuous for the operator norm of $\mathcal{K}$, then $\mathcal{T}$ is said to be \textbf{uniformly
continuous}, which is equivalent to the following assertion \cite{bratteli}: There exists a bounded linear operator $L$ on $\mathcal{K}$ such that
  $T_t=e^{tL},\;\forall t\geq 0.$
In this case,
\begin{equation}\label{defi-gerlim}\nonumber
L=\lim_{t\rightarrow\infty}\frac{1}{t}(T_t-I_\mathcal{K})
\end{equation}
and the operator $L$ is called the \textbf{generator} of $\mathcal{T}.$ \index{generator of $\mathcal{T}$}

A semigroup $\mathcal{T}:=(\mathcal{T}_t)_{t\geq 0}$ of completely positive (CP) and trace-preserving (TP) maps acting on the set of trace-class operators on $\mathcal{K},$ denoted $\mathcal{I}_1(\mathcal{K}),$ is called a \textbf{Quantum Markov Semigroup} (QMS) on  $\mathcal{I}_1(\mathcal{K}).$ Assuming $\lim_{t\rightarrow 0}||\mathcal{T}_t- I||=0,$ $\mathcal{T}$ has a generator $\mathcal{L}=\lim_{t\rightarrow 0^+}(\mathcal{T}_t-I)/t$ (see \cite{Lind}), which is a bounded operator on $\mathcal{I}_1(\mathcal{K}),$ also called a \textbf{Lindblad operator}. {\color{black} Following the convention of \cite{bardet}, we defined the Lindblad generator directly on trace-class operators, which is sufficient for our purposes as we work exclusively with density operators and their evolution. This approach avoids the technicalities of $*$-algebra domains and focuses on the physically relevant states.}

\begin{defi}
Let $V$ be a finite or countable infinite set and $\mathcal{H}$ be a Hilbert space of the form \eqref{Hdef}. A \textbf{Continuous-time Open Quantum Walk} in $V$ is an uniformly continuous QMS on $\mathcal{I}_1(\mathcal{H})$ with Lindblad
operator of the form
\begin{equation} \label{eq-def}
\begin{split}
\mathcal{\mathcal{L}}:\mathcal{I}_1(\mathcal{H}) &\rightarrow   \mathcal{I}_1(\mathcal{H})\\ 
  \rho &\mapsto  -i[\textbf{H},\rho]+\sum_{i,j\in V}\left(S_i^j\rho S_i^{j^*}-\frac{1}{2}\{S_i^{j*}S_i^j,\rho\}\right),
\end{split}
\end{equation}
where $[A,B]\equiv AB-BA$ and $\{A,B\}\equiv AB+BA.$

To rearrange Equation \eqref{eq-def},
we can write $S_i^j=R_i^j\otimes\ket{j}\bra{i}$ for bounded operators $R_i^j:\mathfrak{h}_i\to\mathfrak{h}_j,$ the bounded operators $\textbf{H}$ and $S_i^j$ on $\mathcal{H}$ are of the form $\textbf{H}=\sum_{i\in V}H_i\otimes\ket{i}\bra{i},$ $H_i$ is self-adjoint on $\mathfrak{h}_i,$ and $\sum_{i,j\in V}S_i^{j*}S_i^j$ converges in the strong sense.
\end{defi}

Letting $\rho=\sum_{i\in V}\rho(i)\otimes\ket{i}\bra{i}\in
\mathcal{S}(\mathcal{H}\otimes\mathcal{K}),$ denoting $$e^{t\mathcal{L}}(\rho)=\mathcal{T}_t(\rho)=\sum_{i\in V}\rho_t(i)\otimes\ket{i}\bra{i},\quad\forall t\geq 0,$$ we have
$$
\frac{d}{dt}\rho_t(i)=-i[H_i,\rho_t(i)]+\sum_{j\in V}\left(R_j^i\rho_t(j) R_j^{i^*}-\frac{1}{2}\{R_i^{j*}R_i^j,\rho_t(i)\}\right).
$$

Equation \eqref{eq-def} can be written as given in \cite[Equation 18.7]{bardet}:
\begin{equation}\label{alternativeLindblad}
\mathcal{L}(\rho)=\sum_{i\in V}\left(G_i\rho(i)+\rho(i)G_i^*+\sum_{j\in V}R_j^i\rho(j) R_j^{i*}\right)\otimes\ket{i}\bra{i},\qquad
G_i=-iH_i-\frac{1}{2}\sum_{j\in V}R_i^{j*}R_i^j.
\end{equation}

In this paper we deal with the quantum trajectories of CTOQWs, which are more complex than the trajectories of discrete-time HOQWs. Thus, we recall that for a given finite or countably infinite set of vertices $V,$ a CTOQW is a stochastic process evolving on a Hilbert space of the form \eqref{Hdef}, the label $i\in V$ represents the position of the walker and when the walker is located at $i\in V,$ its internal state is encoded in $\mathfrak{h}_i,$ that is, $\mathfrak{h}_i$  describes the internal degrees of freedom of the walker when it is at site $i\in V.$

Let $i\in V$ and $\rho\in\mathcal{D}(\mathfrak{h}_i).$ Starting the walk at site $\ket{i}$ with initial density operator $\rho,$ the quantum measurement of the ``position" gives rise to a probability distribution $p_0$ on $V,$ such that
$$
p_0(j)=\mathbb{P}(\mbox{the quantum particle is in site}\;\ket{j})=\mathrm{Tr}(\rho(j)),
$$
and for evolution at time $t\geq 0,$
\begin{equation}\label{rho_t(i)}
p_t(j)=\mathbb{P}(\mbox{the quantum particle, at time }t,\mbox{ is in site}\;\ket{j})=\mathrm{Tr}(\rho_{t}(j)),
\end{equation}
where
$$
e^{t\mathcal{L}}(\rho)=\sum_{k\in V}\rho_t(i)\otimes\ket{k}\bra{k}.
$$

\subsection{Quantum Trajectories}

Let $(\Omega,\mathcal{F},(\mathcal{F}_t)_{t\geq 0},\mathbb{P})$ be a probability space where independent Poisson point processes $N^{ij},i,j\in V,\; i\neq j$ ($N^{ii}=0$ by convention) on $\mathbb{R}^2$ are defined. The jump from site $i$ to site $j$ on the graph $V$ will be governed by these Poisson point processes \cite{bardet}:

\begin{defi}\label{XtDefi}
Consider a CTOQW with generator of the form \eqref{alternativeLindblad} and an initial density operator $\mu=\sum_{i\in V}\rho(i)\otimes\ket{i}\bra{i}\in\mathcal{S}(\mathcal{H}\otimes\mathcal{K}).$ The quantum trajectory describing the indirect measurement of the position of the CTOQW is the Markov chain represented by the density operators $(\mu_t)_{t\geq 0}$ such that $\mu_0=\rho_0\otimes\ket{X_0}\bra{X_0},$ where $X_0$ and $\rho_0$ are random variables with distribution
$$
\mathbb{P}\left((X_0,\rho)=\left(i,\frac{\rho(i)}{\mathrm{Tr}(\rho(i))}\right)\right)=\mathrm{Tr}\left(\rho(i)\right)\mbox{ for all }i\in V,
$$
and $\mu_t=:\rho_t\otimes\ket{X_t}\bra{X_t}$ satisfies the stochastic differential equation
\begin{equation}\label{muTrajectory}
\begin{split}
   \mu_t =& \mu_0+\int_{0}^{t}M(\mu_{s^-})ds \\
     +& \sum_{ij}\int_{0}^{t}\int_{\mathbb{R}}\left(\frac{S_{i}^{j}\mu_{s^-}S_{i}^{j*}}{\mathrm{Tr}(S_{i}^{j*}\mu_{s^-}S_{i}^{j*})}-\mu_{s^-}\right)1_{0<y<\mathrm{Tr}(S_{i}^{j}\mu_{s^-}S_{i}^{j*})}N^{ij}(dy,ds)
\end{split}
\end{equation}
for all $t\geq 0,$ where
$$
M(u)=\mathcal{L}(u)-\sum_{ij}\left(S_{i}^{j}\mu S_{i}^{j*}-\mu\mathrm{Tr}(S_{i}^{j}\mu S_{i}^{j*})\right).
$$

Hence, for a fixed $\mu=\sum_{i}\rho(i)\otimes\ket{i}\bra{i}\in\mathcal{S}(\mathcal{H}\otimes\mathcal{K}),$
$$
M(\mu)=\sum_{i}\left(G_i\rho(i)+\rho(i)G_i^*-\rho(i)\mathrm{Tr}\left(G_i\rho(i)+\rho(i)G_i^*\right)\right)\otimes\ket{i}\bra{i}.
$$
\end{defi}

For simplicity, suppose $X_0=i_0$ for some $i_0\in V$ and $\rho_0\in V(\mathfrak{h}_{i_0}).$ For all $t\geq 0,$ consider the solution of Equation \eqref{muTrajectory}
$$
\eta_t=\rho_0+\int_{0}^{t}\left(G_{i_0}\eta_s+\eta_sG_{i_0}^*-\eta_s\mathrm{Tr}\left(G_{i_0}\eta_s+\eta_sG_{i_0}^*\right)\right)ds,
$$
which is a density operator on $\mathfrak{h}_{i_0}.$ For $j\neq i_0,$ define
$$
T_1^j=\mathrm{inf}\left\{t\geq 0;N^{i_0,j}\left({u,y|0\leq u\leq t,0\leq y\leq \mathrm{Tr}(R_{i_0}^{j}\eta_u R_{i_0}^{j*})}\right)\geq 1\right\}.
$$
Since the random variables $T_1^j$ are mutually independent and nonatomic, we can define $T_1=\inf_{j\neq i_0}\{T_1^j\}$ once there exists a unique $j\in V$ such that $T_1^j=T_1.$ The random variable $T_1$ is said to be the \textbf{first jump time} of the CTOQW conditional on $X_0=i_0.$

The first jump time to site $\ket{j}$ is then denoted by $T_1^j$ and has distribution
$$
\mathbb{P}(T_1^j>\varepsilon)=e^{-\int_0^\varepsilon\mathrm{Tr}(R_{i_0}^{j}\eta_uR_{i_0}^{j*})du},
$$
thus
$$
\mathbb{P}(T_1\leq\varepsilon)\leq\varepsilon\sum_{j\neq i_0}\|R_{i_0}^{j*}R_{i_0}^{j}\|.
$$
The strong convergence of $\sum_{ij}S_{i}^{j*}S_{i}^{j}$ implies that $\mathbb{P}(T_1>0)=1.$ Thereby, on $[0,T_1],$ we can define the solution $(X_t,\rho_t)_{t\geq 0}$ as
\begin{eqnarray}
\nonumber (X_t,\rho_t)  &=& (i_0,\eta_t)\mbox{ for }t\in[0,T_1)\mbox{ and } \\
\nonumber (X_{T_1},\rho_{T_1})  &=& \left(j,\frac{R_{i}^{j}\eta_{T_1-}R_{i}^{j*}}{\mathrm{Tr}(R_{i}^{j}\eta_{T_1-}R_{i}^{j*})}\right)\;\mbox{ if }T_1=T_1^j.
\end{eqnarray}

Now we solve
\begin{equation*}\label{eta_t defi}
\eta_t=\rho_{T_1}+\int_{0}^{t}\left(G_{j}\eta_s+\eta_sG_{j}^*-\eta_s\mathrm{Tr}\left(G_{j}\eta_s+\eta_sG_{j}^*\right)\right)ds
\end{equation*}
and then obtain the second jump time $T_2.$ And so on we obtain an increasing sequence of jumps $(T_n)_n$ with $\lim_{n\rightarrow\infty}T_n=\infty$ almost surely (see section 18.2.3 of \cite{bardet} for more details). This means that the walk does not explode, thus the walker has a finite number of jumps in a finite interval. For details concerning explosions of classical Markov chains, see \cite[Section 2.2]{norris}.

\subsection{Recurrence of CTOQWs}

Moving forward, we deal with the following class of 2-dimensional CTOQWs, that is, we will consider the nearest-neighbor quantum random walk on $V=\mathbb{Z}^2.$

\begin{defi}
Consider a CTOQW on $V=\mathbb{Z}^2$.
If there exist $A_1,A_2,A_3,A_4,H\in\mathcal{B}(\mathbb{C}^n)$ such that
$$
H_k=H\;\forall k\in \mathbb{Z}^2,\quad
A_1=
R_{(i,j)}^{(i+1,j)},\;
A_2=
R_{(i,j)}^{(i,j+1)},\;
A_3=
R_{(i,j)}^{(i-1,j)},\;
A_4=
R_{(i,j)}^{(i,j-1)},
$$
and $R_{x}^{y}=0_n$ for the remaining $(x,y)\in \mathbb{Z}^2,$
then the CTOQW is said to be \textbf{induced by a coin} $(A,H)$ of dimension $n.$
\end{defi}

For this kind of CTOQW we can let $\mathfrak{h}_i=\mathfrak{h}$ for all $i\in \mathbb{Z}^2.$ If $\dim(\mathfrak{h})=n,$ then $A_1,A_2,A_3,A_4$ and $H$ can be represented by square matrices of order $n.$

The \textbf{auxiliary map} of a coin $(A,H)$ is defined as
$$
\mathbb{L}(\rho)=G\rho+\rho G^*+\sum_{j=1}^4A_j\rho A_j^*,\qquad
G=-iH-\frac{1}{2}\sum_{k=1}^4A_k^*A_k.
$$
The map $\mathbb{L}$ has at least one stationary state $\rho_\infty,$ that is, $\mathbb{L}(\rho_\infty)=0$ \cite{baum}. When the stationary state is unique, we define
$$m=\begin{bmatrix}
m_1\\m_2
\end{bmatrix},\quad
m_1=\Tr\left(A_1\rho_\infty A_1^*\right)-\Tr\left(A_3\rho_\infty A_3^*\right),\quad
m_2=\Tr\left(A_2\rho_\infty A_2^*\right)-\Tr\left(A_4\rho_\infty A_4^*\right).
$$

To demonstrate the recurrence criteria of this work for CTOQWs, we recall the following limit theorems for such walks.

\begin{teo}{\color{black}\cite[Proposition 3.0.9]{bri}}\label{LazyCLT} Consider a coin $(A,H)$, where its auxiliary map has a stationary state $\rho_\infty.$ Then
\begin{equation*}\label{eq-CLT}
\frac{X_t-tm}{\sqrt{t}}\stackrel{t\rightarrow\infty}{\longrightarrow}N(0,{\color{black}\Sigma})
\end{equation*}
for some {\color{black}$\Sigma\in \mathbb{M}_2(\mathbb{C}).$}
\end{teo}

\begin{prop}[Law of Large Numbers for CTOQWs ]\cite[Theorem 15]{loebens}\label{PropLawLargeNumbers} 
\label{LawofLargeNumbersCT}
If the auxiliary map of a coin $(A,H)$ has a unique stationary state, then
\begin{equation*}\label{LLN}
\frac{X_t}{t}\rightarrow m\quad \mbox{a.s}.
\end{equation*}

\end{prop}

Let us recall the classical definition of recurrence of CTOQWs. For this, $p_{ji;\rho}(t)$ will denote the probability of being at site $j$ at time $t,$ given that we started at site $i,$ with initial density $\rho$ concentrated at $i$:
\begin{equation*}
p_{ji;\rho}(t)=p_{t}(\rho\otimes |i\rangle \to |j\rangle)=\mathrm{\mathrm{Tr}}(\rho_t(j)\otimes\ket{j}\bra{j})=\mathrm{\mathrm{Tr}}\left(e^{t\mathcal{L}}(\rho\otimes\ket{i}\bra{i})(I\otimes\ket{j}\bra{j})\right).
\end{equation*}
Therefore, the walk starts with a density operator $\rho$ concentrated at some vertex $\ket{i},$ takes the evolution up to time $t> 0$ through the quantum Markov semigroup generated by $\mathcal{L},$ producing a new density operator
$$\rho_t=\sum_k\rho_t(k)\otimes\ket{k}\bra{k}=e^{t\mathcal{L}}(\rho\otimes\ket{i}\bra{i}),\quad \mathrm{Tr}\left(\sum_k\rho_t(k)\right)=1.$$
Now we project $\rho_t$ onto the subspace generated by vertex $\ket{j}$ giving
$$
e^{t\mathcal{L}}(\rho\otimes\ket{i}\bra{i})(I\otimes\ket{j}\bra{j})=\rho_t(j)\otimes\ket{j}\bra{j},
$$
which represents the data concentrated at vertex $\ket{j}$ at time $t.$

Let $i\in V,\rho\in \mathcal{D}(\mathfrak{h}_i).$ We say that vertex $i$ is
\begin{itemize}
  \item $\rho$-\textbf{recurrent}\index{$\rho$-recurrent vertex} if
  \begin{equation*}
    \int_{0}^{\infty}p_{ii;\rho}(t)dt=\infty.
  \end{equation*}
Otherwise, $i$ is said to be $\rho$-\textbf{transient};
  \item \textbf{recurrent},\index{recurrent vertex} if $i$ is $\rho-$recurrent for all $\rho\in \mathcal{D}(\mathfrak{h}_i);$
  \item \textbf{transient},\index{transient vertex} if $i$ is $\rho-$transient for all $\rho\in \mathcal{D}(\mathfrak{h}_i).$
\end{itemize}

By definition, $i$ is $\rho$-recurrent when the mean number of returns to $i$ is infinite. 
\begin{defi}
A CTOQW is said to be:
\begin{itemize}
  \item \textbf{recurrent}\index{recurrent CTOQW} if every vertex is recurrent;
  \item \textbf{transient}\index{transient CTOQW} if every vertex is transient.
\end{itemize}
\end{defi}

Within this setting, recurrence can be naturally associated with quantum trajectories. This connection relies on the homogeneity of the coin, which guarantees the validity of the result stated in the next proposition for a CTOQW induced by a coin.

\begin{prop}\label{homo}
A CTOQW induced by a coin is $\rho$-recurrent if and only if it is $\rho_{T_n}$-recurrent for any possible jump time $T_n.$\end{prop}
\begin{proof}Note that if we start the walk with $(X_0,\rho_0)=(i,\rho),$ then at some time $t>0,$ we have $(X_t,\rho_t)=(j,\rho_T).$ If we refresh the walk at this instant, then the walk is still recurrent with respect to the density $\rho_t,$ by the homogeneity in space. The other hand follows the same idea, assuming by contraposition that the walk is $\rho$-transient.
\end{proof}

\begin{prop}\label{prop_1enclosure}
If $(A,H)$ is $\tau$-recurrent for some density $\tau\in\mathcal{D}(\mathfrak{h})$ and $\mathfrak{h}$ has at most one non-trivial invariant subspace for
$\Phi:\mathfrak{h}\rightarrow \mathfrak{h}$ given by
$$
\Phi(.)=\sum_k A_k(.)A_k^*,
$$
then $(A,H)$ is recurrent. {\color{black} In particular, if $(A,H)$ is $\tau$-recurrent for some density $\tau\in\mathcal{D}(\mathfrak{h})$ and $\mathfrak{h}$ has at most one minimal enclosure for
$\Phi,$ then $(A,H)$ is recurrent.}
\end{prop}
\begin{proof}
Let us start supposing that $(A,H)$ is recurrent with respect to some density operator $\tau$ with support on $\mathcal{V}.$ If $(A,H)$ starts the walk with initial density operator $\rho\in \mathcal{D}(\mathfrak{h}),$ then $\rho_{T_n}$ is of the form
\begin{equation}\label{rhon}
\rho_{T_n}=\frac{A_k\eta A_k^*}{\Tr \left(A_k\eta A_k^*\right)}\text{ for some }k\in\{1,2,3,4\},
\end{equation}
{\color{black}where supp$(\eta)\cap\mathcal{V}\neq\{0\},$ thus $\mathcal{V}\subset\text{supp}(\rho_{T_n})$ for some $n\geq 0,$ otherwise there would be another non-trivial invariant subspace for $\Phi.$ Equation \eqref{rhon} allows us to find some $a>0$ for this $n$ such that
$$
\sigma:=\rho_{T_n}-\frac{\tau}{a}\geq 0.
$$}

Therefore,
$$
\frac{1}{\Tr(\sigma)}\int_0^\infty p_{ii;\rho_{T_n}}(t)\;\text{d}t=
\int_0^\infty p_{ii;\frac{\sigma}{\Tr(\sigma)}}(t)\;\text{d}t+
\frac{1}{a\Tr(\sigma)}\int_0^\infty p_{ii;\tau}(t)\;\text{d}t=\infty.
$$
By Proposition \ref{homo}, $(A,H)$ is $\rho$-recurrent. {\color{black} The particular case holds, since $\Phi$ has only one non-trivial invariant subspace for $A_k, \;k=1,2,3,4,$ when there is only one minimal enclosure. Indeed, the invariant subspaces of $\Phi$ correspond to enclosures, since the
projections in the fixed point set are exactly the projections onto
enclosures (see \cite[Corollary 2.2.6]{girotti-thesis}). Moreover, space $\mathcal{R}$
admits a decomposition in minimal enclosures, as in Equation \eqref{R_grande},
where every non-trivial enclosure
contains a minimal enclosure. Hence, if $\mathfrak h$ has at most one minimal
enclosure for $\Phi$, then $\Phi$ has at most one non-trivial invariant subspace,
and the Proposition applies.
}
\end{proof}

\subsection{Recurrence Criterion for CTOQWs on the Grid}

This section is devoted to exhibit a recurrence criterion for finite dimensional coins $(A,H),$ where the auxiliary map has at most one common invariant subspace. The result uses a CLT for CTOQWs and the following Lemma.
\begin{lema}\label{lemaMax}Let us consider a homogeneous  CTOQW on $\mathbb{Z}^d.$ Let $\varepsilon>0$ and $m\geq 1$ be an integer. Then, for every $\rho \in \mathcal{D}(\mathfrak{h})$,
$$
\int_{0}^\infty \mathbb{P}_{0,\rho}(|X_t|<m\varepsilon) dt
\leq
 (2m)^d\cdot \max_{\sigma \in \mathcal{D}}(\mathfrak{h})\int_{0}^\infty\mathbb{P}_{0,\sigma}(|X_t|<\varepsilon)dt.
$$
\end{lema} 

\textbf{Proof.} Firstly, note that
\beq\label{eqlemchung}
\int_{0}^\infty \mathbb{P}_{0,\rho}\left(|X_t|<m.\varepsilon\right)dt \leq 
\int_{0}^\infty \sum_{k}\mathbb{P}_{0,\rho}\left(X_t\in k.\varepsilon+[0,\varepsilon)^d\right)dt,
\eeq
where the sum runs over $k$ in the set $\{-m,-m+1,\ldots,m-2,m-1\}^d$.

Let 
$$T_k = \inf \left\{ z \geq 0 : X_z\in k.\varepsilon+\left[0,\varepsilon\right)^d \right\}\;\text{ and }\;\zeta_\rho^{k,\varepsilon} = \int_{0}^\infty \mathbb{P}_{0,\rho}\left  (X_t\in k.\varepsilon +[0,\varepsilon)^d\right)dt.$$. 
By the law of total probability and Fubini's theorem, we can rewrite $\zeta_\rho^{k,\varepsilon}$ as
\begin{equation}
\begin{split}
\zeta_\rho^{k,\varepsilon}
=&\int_{0}^\infty\int_{0}^t \mathbb{P}_{0,\rho}\left  (X_t\in k.\varepsilon +[0,\varepsilon)^d,\;T_k=y\right)dy\;dt\\
=&
\int_{0}^\infty \int_{0}^\infty \mathbb{P}_{0,\rho}(|X_t - X_y |<\varepsilon,T_k=y)dy\; dt\\
=&\int_{0}^\infty \int_{y}^\infty \mathbb{P}_{0,\rho}(|X_t - X_y |<\varepsilon,T_k=y)dt\;dz\\
=&
\int_{0}^\infty \mathbb{P}_{0,\rho}(T_k=z)
\int_{y}^\infty \mathbb{P}_{0,\rho}(|X_t - X_y |<\varepsilon)dt\;dy,
\end{split}
\end{equation}
where the last equality follows from the independency between the events $\{T_k=y\}$ and$\{|X_t-X_y|<\varepsilon \}.$

Denote $\mathbb{P}_{0,\rho}^{k,y}=\mathbb{P}_{0,\rho}\left(T_k=y\right).$  An application of Fubini's theorem gives
\begin{equation}\nonumber
\begin{split}
\zeta_{\varepsilon,k}^\rho=& \int_{0}^{\infty}\mathbb{P}_{0,\rho}^{k,y}\int_{y}^{\infty}\mathbb{P}_{0,\rho}\left(|X_t-X_y|<\varepsilon\right)dtdy \\
     =&\int_{0}^{\infty}\mathbb{P}_{0,\rho}^{k,y}\int_{y}^{\infty}\sum_{\sigma\in\mathcal{D}(\mathfrak{h}),j\in\mathbb{Z}}\mathbb{P}_{0,\rho}\left(|X_t-X_y|<\varepsilon,(X_y,\rho_y)=(j,\sigma)\right)dtdy \\
     =&\int_{0}^{\infty}\mathbb{P}_{0,\rho}^{k,y}\sum_{\sigma\in\mathcal{D}(\mathfrak{h}),j\in\mathbb{Z}}\int_{y}^{\infty}\mathbb{P}_{0,\rho}\left(|X_t-X_y|<\varepsilon|(X_y,\rho_y)=(j,\sigma)\right)\mathbb{P}_{0,\rho}\left((X_y,\rho_y)=(j,\sigma)\right)dtdy.
\end{split}
\end{equation}

Due to the spatial homogeneity of the walk and the change of variables $s=t-y,$ we have
\begin{equation}\nonumber
\begin{split}
\zeta^\rho_{\varepsilon,k}=&\int_{0}^{\infty}\mathbb{P}_{0,\rho}^{k,y}\sum_{\sigma\in\mathcal{D}(\mathfrak{h}),j\in\mathbb{Z}}
\int_{y}^{\infty}\mathbb{P}_{j,\sigma}\left(|X_{t-y}-X_0|<\varepsilon\right)dt\mathbb{P}_{0,\rho}\left((X_y,\rho_y)=(j,\sigma)\right)dy \\
=&\int_{0}^{\infty}\mathbb{P}_{0,\rho}^{k,y}\sum_{\sigma\in\mathcal{D}(\mathfrak{h}),j\in\mathbb{Z}}\int_{0}^{\infty}\mathbb{P}_{j,\sigma}
\left(|X_{s}-j|<\varepsilon\right)ds\mathbb{P}_{0,\rho}\left((X_y,\rho_y)=(j,\sigma)\right)dy \\
=&\int_{0}^{\infty}\mathbb{P}_{0,\rho}^{k,y}\sum_{\sigma\in\mathcal{D}}(\mathfrak{h})\int_{0}^{\infty}\mathbb{P}_{0,\sigma}
\left(|X_{s}|<\varepsilon\right)ds\sum_{j\in\mathbb{Z}}\mathbb{P}_{0,\rho}\left((X_y,\rho_y)=(j,\sigma)\right)dy.
\end{split}
\end{equation}
The inner integral depends only on the internal state $\sigma$ and, since $\mathcal{D}(\mathfrak{h})$ is compact and the integral is continuous in $\sigma$, it attains a finite maximum. Replacing it by this maximum - which is independent of $\sigma$ and $j$ - yields the estimate:
\begin{equation}\nonumber
\begin{split}
\zeta^\rho_{\varepsilon,k}\leq&\int_{0}^{\infty}\mathbb{P}_{0,\rho}^{k,y}\sum_{\sigma\in\mathcal{D}(\mathfrak{h})}\max_{\tau\in\mathcal{D}(\mathfrak{h})}\int_{0}^{\infty}
\mathbb{P}_{0,\tau}\left(|X_{s}|<\varepsilon\right)ds\sum_{j\in\mathbb{Z}}\mathbb{P}_{0,\rho}\left((X_y,\rho_y)=(j,\sigma)\right)dy\\
=&\int_{0}^{\infty}\mathbb{P}_{0,\rho}^{k,y}\max_{\tau\in\mathcal{D}(\mathfrak{h})}\int_{0}^{\infty}
\mathbb{P}_{0,\tau}\left(|X_{s}|<\varepsilon\right)ds\sum_{\sigma\in\mathcal{D}(\mathfrak{h}),j\in\mathbb{Z}}\mathbb{P}_{0,\rho}\left((X_y,\rho_y)=(j,\sigma)\right)dy\\
=&\int_{0}^{\infty}\mathbb{P}_{0,\rho}^{k,y}dy\max_{\tau\in\mathcal{D}(\mathfrak{h})}\int_{0}^{\infty}
\mathbb{P}_{0,\tau}\left(|X_{s}|<\varepsilon\right)ds\\
=&\max_{\tau\in\mathcal{D}(\mathfrak{h})}\int_{0}^{\infty}
\mathbb{P}_{0,\tau}\left(|X_{s}|<\varepsilon\right)ds.
\end{split}
\end{equation}

Therefore, Equation \eqref{eqlemchung} and the inequality above give
$$
\int_{0}^\infty \mathbb{P}_{0,\rho}(|X_t|<m\varepsilon)dt \leq 
 \sum_{k}
\zeta_\rho^{k,\varepsilon}
\leq
 \sum_{k}
\max_{\tau \in \mathcal{D}(\mathfrak{h})}
\int_{0}^\infty 
\mathbb{P}_{0,\tau}(|X_{s} |<\varepsilon)ds
=
(2m)^d \cdot
\max_{\tau \in \mathcal{D}(\mathfrak{h})}
\int_{0}^\infty 
\mathbb{P}_{0,\tau}(|X_{s} |<\varepsilon)ds.
$$

\begin{teo}\label{TeoGauss}
If $X_t$ denotes a CTOQW on $\mathbb{Z}^2$ and $X_t/t^{1/2}\Rightarrow$ a Gaussian measure, then the CTOQW is recurrent for some density.
\end{teo}

\begin{proof}
Let us denote $u(t,m)=\mathbb{P}(|X_t|<m).$ By Lemma \ref{lemaMax},
$$
\max_{\tau\in D}\int_{0}^\infty \mathbb{P}(|X_t|<1)dt=
\max_{\tau\in D}\int_{0}^\infty u(t,1)dt\geq (4m^2)^{-1}
\int_{0}^\infty u(t,m)dt.
$$

Assume that $m/\sqrt{t}\rightarrow c,$ then
\begin{equation}\label{UInt}
u(t,m)\rightarrow \int_{[-c,c]^2}f(x)dx,
\end{equation}
where $f(x)$ represents the density of the limiting normal distribution. We denote
the right-hand side of Equation \eqref{UInt} by $g(c)$ and let $t=\theta m^2$ to obtain
$$
u(\theta m^2,m)\rightarrow g(\theta^{-1/2}).
$$
Now we use the change of variables
$$
m^{-2}\int_{0}^\infty u(t,m)dt=\int_0^\infty
u(\theta m^2,m)d\theta,
$$
let $m\rightarrow\infty,$ and use Fatou's lemma to get
\begin{equation}\label{fatou}
\liminf_{m\rightarrow\infty}(4m^2)^{-1}\int_{0}^\infty u(t,m)dt\geq 4^{-1}\int_{0}^\infty g(\theta^{-1/2})d\theta.
\end{equation}

Since the normal density is positive and continuous at $0,$
$$
g(c)=\int_{[-c,c]^2}n(x)dx\sim n(0)(2c)^2
$$
as $c\rightarrow 0.$ So $g(\theta^{-1/2})\sim 4n(0)/\theta$ as $\theta\rightarrow \infty,$ the integral in  
\eqref{fatou} diverges. As a result, $\max_{\tau\in D}\int_{0}^\infty \mathbb{P}(|X_t|<1)dt=\infty,$ leading us to conclude that the walk is $\tau$-recurrent.
\end{proof}

Note that the maximum in Lemma \ref{lemaMax} can be assumed to be taken over the jump times $(T_n)_n.$ Therefore, {\color{black}by Proposition \ref{prop_1enclosure}} and Theorem \ref{TeoGauss} we obtain the following result.

\begin{teo}\label{TeoGauss2}
If $X_t$ denotes a CTOQW induced by a coin $(A,H),$ there is at most one minimal enclosure for $\Phi$ and $X_t/t^{1/2}\Rightarrow$ a Gaussian measure, then the CTOQW is recurrent.
\end{teo}

We already have a sufficient condition to recurrence given by the Theorem  \ref{TeoGauss2}. For transience, we also have a sufficient condition:
\begin{teo}\label{tran}
Let us consider a CTOQW induced by a coin $(A,H),$ having at most one minimal enclosure, and let $\rho_\infty$ be the unique stationary state of the auxiliary map. If $m\neq 0,$ then the  CTOQW is transient.
\begin{proof}
Let us start the CTOQW at some vertex $\ket{i}\in \mathbb{Z}^2$ with initial density operator $\rho.$ By the Law of Large Numbers, Proposition \ref{PropLawLargeNumbers},
$$
\frac{X_t}{t}\rightarrow m\neq 0^2.
$$
This means that at least one coordinate of $X_t$ is unbounded, from where we conclude that $$\mathbb{P}\left(X_t=X_0 \;\text{i.o.}\right)=0.$$ Hence, the CTOQW is transient, {\color{black}since this is valid for any $\rho.$}
\end{proof}
\end{teo}

Finally, we can present a recurrence criteria for  CTOQWs on $\mathbb{Z}^2$ induced by a coin $(A,H).$

\begin{teo}[Recurrence criteria for CTOQWs on $\mathbb{Z}^2$]\label{ct.criterion}
Let us consider a CTOQW induced by a coin $(A,H),$ having at most one minimal enclosure, and let $\rho_\infty$ be the unique stationary state of the auxiliary map. Then

\medskip

\begin{itemize}
\item $m=0\;\Rightarrow\;$ the CTOQW is recurrent; 

\medskip

\item $m\neq 0\;\Rightarrow\;$ the CTOQW is transient.
\end{itemize}

\begin{proof} For $m\neq 0,$ the result is given in Theorem \ref{tran}. It remains to prove that $m=0$ implies that the CTOQW is recurrent. If $m=0,$ then by Theorem \ref{LazyCLT}
$$
\frac{X_t}{\sqrt{t}}=
\frac{X_t-t.m}{\sqrt{t}}\Rightarrow \text{ a non-degenerate normal distribution,}
$$
thus we use Theorem \ref{TeoGauss} to obtain a density operator $\tau$ such that the CTOQW is $\tau$-recurrent. Therefore, $(A,H)$ is recurrent by Theorem \ref{TeoGauss2}.
\end{proof}
\end{teo}

\subsection{Recurrence Criterion for HOQWs on the Grid}

We begin with an easily verifiable criterion based on the quantum jump chains associated with CTOQWs, as presented in \cite{quantum.jump}. Let $\Pi$ denote a discrete HOQW given by
$$\Pi(\rho)=\sum_{i,j\in V}M_j^i\rho(j)M_j^{i*}\otimes\ket{i}\bra{i}.$$ This HOQW is the open quantum jump chain of the CTOQW with Lindblad generator
\begin{equation}\label{LPhiIg}
\mathbb{L}=\Psi-I,\;\;\Psi(\rho)=\sum_{i,j\in V}B_j^i\rho(j)B_j^{i*},\;B_j^i=M_j^i\otimes\ket{i}\bra{j},
\end{equation}
where $I$ is the identity operator acting on $\mathcal{H}$ \cite{quantum.jump}.

\begin{teo}\cite[Theorem 5.6]{quantum.jump}\label{d.vs.c} A vertex $i\in V$ is $\rho$-recurrent for the HOQW $\Pi$ if and only if it is $\rho$-recurrent for the CTOQW generated by $\mathbb{L}$ given by Equation \eqref{LPhiIg}.
\end{teo}

{\color{black}Theorem \ref{d.vs.c} relies on the open quantum jump chain construction introduced in \cite{quantum.jump}. Given a CTOQW with Lindblad generator $\mathbb{L}$, one defines a discrete-time OQW $\Phi$ via $\Phi = \Psi - I$, where $\Psi(\rho) = \sum_{i,j} B_j^i \rho(j) B_j^{i*}$ and $B_j^i = M_j^i \otimes |i\rangle\langle j|$ are the jump operators. This construction yields a discrete-time quantum walk that records the sequence of jump events and the post-jump quantum states. The preservation of recurrence follows from the fact that the jump operators encode the same transitions as the continuous-time process. This equivalence holds because the jump times in the continuous-time process are almost surely finite and the sequence of visited sites and internal states coincides with that of the discrete-time walk. For a detailed exposition of this construction and the proof of recurrence equivalence, we refer the reader to \cite[Section 5]{quantum.jump}}.

We consider an HOQW with a set of vertices $V=\mathbb{Z}^2.$ Analogously to the continuous-time version, if there exists $D_1,D_2,D_3,D_4\in\mathcal{B}(\mathbb{C}^n)$ such that
$$
D_1=
M_{(i,j)}^{(i+1,j)},\;
D_2=
M_{(i,j)}^{(i,j+1)},\;
D_3=
M_{(i,j)}^{(i-1,j)},\;
D_4=
M_{(i,j)}^{(i,j-1)},
$$
and $M_{x}^{y}=0_n$ for the remaining $(x,y)\in \mathbb{Z}^2,$
then the HOQW $\Pi$ is said to be \textbf{induced by a coin} $(D)$ of dimension $n.$ 

The HOQW induced by a coin $(D)$ is the homogeneous, nearest-neighbor HOQW on the grid, with the right, up, left and down transitions given by $D_1,D_2,D_3,D_4,$ respectively. Therefore, the \textbf{quantum trajectory} of those walks, starting from a state $\tau$ of the form $\tau=\sum_{i \in \mathbb{Z}}\rho(i) \otimes |i\rangle\langle i|,$
is any path generated by the Markov chain $(X_n,\tau_n)_{n\geq 0}$, where $X_n$ denotes the position of the particle at time $n$ and $\tau_n$ its internal degree. The transition probabilities are given by
\begin{equation}\nonumber
\begin{split}
\mathbb{P}\left(\, ( X_{n+1},\tau_{n+1}) = \left((i+1,j), \frac{D_1\sigma D_1^*}{\mathrm{Tr}(D_1\sigma D_1^*)}\right) \; \Bigg| \; (X_n,\rho_n) = ((i,j), \sigma)\,\right)& = \mathrm{Tr}(D_1\sigma D_1^*),\\
\mathbb{P}\left(\, ( X_{n+1},\tau_{n+1}) = \left((i,j+1), \frac{D_2\sigma D_2^*}{\mathrm{Tr}(D_2\sigma D_2^*)}\right) \; \Bigg| \; (X_n,\rho_n) = ((i,j), \sigma)\,\right)& = \mathrm{Tr}(D_2\sigma D_2^*),\\
\mathbb{P}\left(\, ( X_{n+1},\tau_{n+1}) = \left((i-1,j), \frac{D_3\sigma D_3^*}{\mathrm{Tr}(D_3\sigma D_3^*)}\right) \; \Bigg| \; (X_n,\rho_n) = ((i,j), \sigma)\,\right)& = \mathrm{Tr}(D_3\sigma D_3^*),\\
\mathbb{P}\left(\, ( X_{n+1},\tau_{n+1}) = \left((i,j-1), \frac{D_4\sigma D_4^*}{\mathrm{Tr}(D_4\sigma D_4^*)}\right) \; \Bigg| \; (X_n,\rho_n) = ((i,j), \sigma)\,\right)& = \mathrm{Tr}(D_4\sigma D_4^*),
\end{split}
\end{equation}
for every $(i,j) \in \mathbb{Z}^2$, $\sigma \in \mathcal{D}(\mathfrak{h})$, and initial law
$$
\mathbb{P}\left(\, (X_{0},\rho_{0}) = \left((i,j), \frac{\rho(i,j)}{\mathrm{Tr}\,\rho(i,j)}\right) \; \right) = \mathrm{Tr}(\rho(i,j)),
$$
and all other transition probabilities are null. The graph of this kind of HOQW is represented in Figure \ref{OQWZ2}.

\begin{figure}[h]
\begin{tikzpicture}
    \def\spacing{2.3}

    \node (v11) at (1*\spacing, 6*\spacing) {};
    \node[circle, draw] (v12) at (2*\spacing, 6*\spacing) {};
    \node[circle, draw] (v13) at (3*\spacing, 6*\spacing) {};
    \node[circle, draw] (v14) at (4*\spacing, 6*\spacing) {};
    \node[circle, draw] (v15) at (5*\spacing, 6*\spacing) {};
    \node (v16) at (6*\spacing, 6*\spacing) {};

    \node[circle, draw] (v21) at (1*\spacing, 5*\spacing) {};
    \node[circle, draw] (v22) at (2*\spacing, 5*\spacing) {};
    \node[circle, draw] (v23) at (3*\spacing, 5*\spacing) {};
    \node[circle, draw] (v24) at (4*\spacing, 5*\spacing) {};
    \node[circle, draw] (v25) at (5*\spacing, 5*\spacing) {};
    \node[circle, draw] (v26) at (6*\spacing, 5*\spacing) {};

    \node[circle, draw] (v31) at (1*\spacing, 4*\spacing) {};
    \node[circle, draw] (v32) at (2*\spacing, 4*\spacing) {};
    \node[circle, draw] (v33) at (3*\spacing, 4*\spacing) {};
    \node[circle, draw] (v34) at (4*\spacing, 4*\spacing) {};
    \node[circle, draw] (v35) at (5*\spacing, 4*\spacing) {};
    \node[circle, draw] (v36) at (6*\spacing, 4*\spacing) {};

    \node[circle, draw] (v41) at (1*\spacing, 3*\spacing) {};
    \node[circle, draw] (v42) at (2*\spacing, 3*\spacing) {};
    \node[circle, draw] (v43) at (3*\spacing, 3*\spacing) {};
    \node[circle, draw] (v44) at (4*\spacing, 3*\spacing) {};
    \node[circle, draw] (v45) at (5*\spacing, 3*\spacing) {};
    \node[circle, draw] (v46) at (6*\spacing, 3*\spacing) {};

    \node[circle, draw] (v51) at (1*\spacing, 2*\spacing) {};
    \node[circle, draw] (v52) at (2*\spacing, 2*\spacing) {};
    \node[circle, draw] (v53) at (3*\spacing, 2*\spacing) {};
    \node[circle, draw] (v54) at (4*\spacing, 2*\spacing) {};
    \node[circle, draw] (v55) at (5*\spacing, 2*\spacing) {};
    \node[circle, draw] (v56) at (6*\spacing, 2*\spacing) {};

    \node (v61) at (1*\spacing, 1*\spacing) {};
    \node[circle, draw] (v62) at (2*\spacing, 1*\spacing) {};
    \node[circle, draw] (v63) at (3*\spacing, 1*\spacing) {};
    \node[circle, draw] (v64) at (4*\spacing, 1*\spacing) {};
    \node[circle, draw] (v65) at (5*\spacing, 1*\spacing) {};
    \node (v66) at (6*\spacing, 1*\spacing) {};

\path[->]          (v32)  edge         [bend left=30]   node[auto] {$D_2$}     (v22);
\path[->]          (v33)  edge         [bend left=30]   node[auto] {$D_2$}     (v23);
\path[->]          (v34)  edge         [bend left=30]   node[auto] {$D_2$}     (v24);
\path[->]          (v35)  edge         [bend left=30]   node[auto] {$D_2$}     (v25);

\path[->]          (v42)  edge         [bend left=30]   node[auto] {$D_2$}     (v32);
\path[->]          (v43)  edge         [bend left=30]   node[auto] {$D_2$}     (v33);
\path[->]          (v44)  edge         [bend left=30]   node[auto] {$D_2$}     (v34);
\path[->]          (v45)  edge         [bend left=30]   node[auto] {$D_2$}     (v35);

\path[->]          (v52)  edge         [bend left=30]   node[auto] {$D_2$}     (v42);
\path[->]          (v53)  edge         [bend left=30]   node[auto] {$D_2$}     (v43);
\path[->]          (v54)  edge         [bend left=30]   node[auto] {$D_2$}     (v44);
\path[->]          (v55)  edge         [bend left=30]   node[auto] {$D_2$}     (v45);

\path[->]          (v22)  edge         [bend left=30]   node[auto] {$D_1$}     (v23);
\path[->]          (v23)  edge         [bend left=30]   node[auto] {$D_1$}     (v24);
\path[->]          (v24)  edge         [bend left=30]   node[auto] {$D_1$}     (v25);

\path[->](v32)edge[bend left=30]node[auto]{$D_1$}(v33);
\path[->](v33)edge[bend left=30]node[auto]{$D_1$}(v34);
\path[->](v34)edge[bend left=30]node[auto]{$D_1$}(v35);

\path[->](v42)edge[bend left=30]node[auto]{$D_1$}(v43);
\path[->](v43)edge[bend left=30]node[auto]{$D_1$}(v44);
\path[->](v44)edge[bend left=30]node[auto]{$D_1$}(v45);

\path[->](v52)edge[bend left=30]node[auto]{$D_1$}(v53);
\path[->](v53)edge[bend left=30]node[auto]{$D_1$}(v54);
\path[->](v54)edge[bend left=30]node[auto]{$D_1$}(v55);

\draw[<-] (v32) -- node[midway, right] {$D_4$} (v22);
\draw[<-] (v33) --node[midway, right] {$D_4$} (v23);
\draw[<-] (v34) --node[midway, right] {$D_4$} (v24);
\draw[<-] (v35) -- node[midway, right] {$D_4$}(v25);

\draw[<-] (v42) --node[midway, right] {$D_4$} (v32);
\draw[<-] (v43) --node[midway, right] {$D_4$} (v33);
\draw[<-] (v44) --node[midway, right] {$D_4$} (v34);
\draw[<-] (v45) --node[midway, right] {$D_4$} (v35);

\draw[<-] (v52) --node[midway, right] {$D_4$} (v42);
\draw[<-] (v53) --node[midway, right] {$D_4$} (v43);
\draw[<-] (v54) -- node[midway, right] {$D_4$}(v44);
\draw[<-] (v55) --node[midway, right] {$D_4$} (v45);

\draw[->] (v23) --node[midway, below] {$D_3$} (v22);
\draw[->] (v24) --node[midway, below] {$D_3$} (v23);
\draw[->] (v25) --node[midway, below] {$D_3$} (v24);

\draw[->] (v33) --node[midway, below] {$D_3$} (v32);
\draw[->] (v34) --node[midway, below] {$D_3$} (v33);
\draw[->] (v35) --node[midway, below] {$D_3$} (v34);

\draw[->] (v43) --node[midway, below] {$D_3$} (v42);
\draw[->] (v44) --node[midway, below] {$D_3$} (v43);
\draw[->] (v45) --node[midway, below] {$D_3$} (v44);

\draw[->] (v53) --node[midway, below] {$D_3$} (v52);
\draw[->] (v54) --node[midway, below] {$D_3$} (v53);
\draw[->] (v55) --node[midway, below] {$D_3$} (v54);

\path (v21) -- node[auto=false]{\large\ldots} (v22);
\path (v31) -- node[auto=false]{\large\ldots} (v32);
\path (v41) -- node[auto=false]{\large\ldots} (v42);
\path (v51) -- node[auto=false]{\large\ldots} (v52);
\path (v22) -- node[auto=false]{\large$\vdots$} (v12);
\path (v23) -- node[auto=false]{\large$\vdots$} (v13);
\path (v24) -- node[auto=false]{\large$\vdots$} (v14);
\path (v25) -- node[auto=false]{\large$\vdots$} (v15);
\path (v25) -- node[auto=false]{\large$\ldots$} (v26);
\path (v35) -- node[auto=false]{\large\ldots} (v36);
\path (v45) -- node[auto=false]{\large\ldots} (v46);
\path (v55) -- node[auto=false]{\large\ldots} (v56);
\path (v62) -- node[auto=false]{\large$\vdots$} (v52);
\path (v63) -- node[auto=false]{\large$\vdots$} (v53);
\path (v64) -- node[auto=false]{\large$\vdots$} (v54);
\path (v65) -- node[auto=false]{\large$\vdots$} (v55);
\end{tikzpicture}
\caption{Homogeneous Open Quantum Walk on $\mathbb{Z}^2$.}
\label{OQWZ2}
\end{figure}

Now we can give a recurrence criteria for HOQWs on $\mathbb{Z}^2.$ To this, note that if $\rho$ is an invariant state for the auxiliary map $\Phi$ of $(D),$ defined as
$$\Phi(.)=\sum_{j=1}^4D_j(.)D_j^*,$$
then it is a stationary state for its open quantum jump chain $\mathbb{L}.$ Consequently, if $\Phi$ has at most one minimal enclosure, then $\Phi$ has a unique invariant state $\rho_\infty$ and we can let
$$m=\begin{bmatrix}
m_1\\m_2
\end{bmatrix},\quad
m_1=\Tr\left(D_1\rho_\infty D_1^*\right)-\Tr\left(D_3\rho_\infty D_3^*\right),\quad
m_2=\Tr\left(D_2\rho_\infty D_2^*\right)-\Tr\left(D_4\rho_\infty D_4^*\right),
$$
to obtain, as a consequence of Theorems \ref{ct.criterion} and \ref{d.vs.c}, the following criterion, {\color{black}which follows the same structure as Theorem \ref{teo_lazy}.}

\begin{teo}[Recurrence criterion for HOQWs on $\mathbb{Z}^2$]\label{OQWcriterion}
Let us consider an HOQW induced by a coin $(D),$ where the map
$$\Phi(.)=\sum_{j=1}^4D_j(.)D_j^*$$
has at most one minimal enclosure, and let $\rho_\infty$ be the unique invariant state of $\Phi.$ Then

\medskip

\begin{itemize}
\item $m=\begin{bmatrix}
    0\\0
\end{bmatrix}\;\Rightarrow\;$ the HOQW is recurrent; 

\medskip

\item $m\neq \begin{bmatrix}
    0\\0
\end{bmatrix}\;\Rightarrow\;$ the HOQW is transient.
\end{itemize}
\end{teo}

{\color{black}The results obtained so far provide recurrence criteria for coins whose associated auxiliary map is ergodic. In the next section, we extend these results to more general settings in which the auxiliary map may admit multiple invariant states. To achieve this, we partition the state space into suitable invariant components, allowing a refined analysis while preserving the role of invariant states in the construction of recurrence criteria. We begin with the case of lazy coins in dimension two, where the low dimensionality enables a complete characterization in terms of eigenvalues and eigenvectors. For higher dimensions, we turn our attention to non-lazy coins.
}

\section{General Recurrence Criteria for HOQWs and CTOQWs}

\subsection{General Recurrence Criterion for HOQWs Induced by a Coin $(L,B,R)$ of Dimension 2}The following theorem is due to \cite{burgath}. 
\begin{teo}\label{adap2}
For a finite quantum channel $\mathcal{L}:\mathcal{B}(\mathfrak{h})\rightarrow\mathcal{B}(\mathfrak{h})$, the following are equivalent:
\begin{description}
\item[i.] $\mathcal{L}$ is ergodic; 
\item[ii.] $\mathcal{L}$ does not have two invariant subspaces $S_1\neq\{0\}$ and $S_2\neq\{0\}$ such that $S_1\cap S_2=\{0\};$
\item[iii.]$\mathcal{L}$ has a minimal invariant subspace, that is, a subspace $S\neq\{0\}$ such that $S\subset S'$ for every invariant subspace $S'\neq \{0\}.$
\end{description}
\end{teo}

Building upon the established preceding theorem, we will advance to the following result, delineating a comprehensive criterion for 2-dimensional coins, extending \cite[Theorem 17]{thomas} to encompass the Lazy HOQW case. This criterion will be structured into two distinct parts: the first adhering to the logic of the previous theorem, and the second articulated in terms of the eigenvalues of the operators of the coin, independent of the invariant densities of the auxiliary map.

\begin{teo}\label{LazyDim2}
Consider an HOQW on $\mathbb{Z}$ induced by a coin $(L,B,R)$ of dimension 2.

(1) If $L,B,R$ have at most one common eigenvector, let $\rho_\infty$ be the unique invariant density of the auxiliary map. Then, we have
$$
(1.1)\quad \textmd{Tr}(L\rho_\infty L^*)\neq\textmd{Tr}(R\rho_\infty R^*)\Rightarrow (L,B,R) \mbox{ is transient},
$$

$$
(1.2)\quad \textmd{Tr}(L\rho_\infty L^*)=\textmd{Tr}(R\rho_\infty R^*)\Rightarrow (L,B,R) \mbox{ is recurrent}.
$$

\bigskip

(2) If $L,B,R$ have two linearly independent eigenvectors in common, $\ket{u_1},\ket{u_2}$ unit, then they satisfy $\ket{u_1}\bot \ket{u_2},$ we can put
$$
L\ket{u_j}=l_j\ket{u_j},\qquad B\ket{u_j}=b_j\ket{u_j},\qquad R\ket{u_j}=r_j\ket{u_j},\qquad j=1,2,
$$
and
\begin{eqnarray}
\nonumber (2.1)\quad |l_1|=|r_1|\mbox{ and }|l_2|=|r_2|&\Rightarrow& (L,B,R)\mbox{ is recurrent,} \\
\nonumber &\\
\nonumber (2.2)\quad |l_1|\neq|r_1|\mbox{ and }|l_2|\neq|r_2|&\Rightarrow& (L,B,R)\mbox{ is transient,} \\
\nonumber &\\
\nonumber (2.3)\quad |l_i|=|r_i|\mbox{ and }|l_j|\neq|r_j|&\Rightarrow& (L,B,R)\mbox{ is transient with respect to }\sigma_j=\ket{u_j}\bra{u_j}\mbox{ and it is recurrent} \\
\nonumber   & & \mbox{ with respect to all densities but }\sigma_j, \mbox{ where }i,j\in\{1,2\},\;i\neq j.
\end{eqnarray}
\begin{proof}
If the elements of the coin do not share at least two orthogonal eigenvectors, then the auxiliary map is ergodic, thus item (1) is a particular case of Theorem \ref{teo_lazy}.

Now we suppose that $L,B,R$ share two unit linearly independent eigenvectors $\ket{u_1},\ket{u_2}.\; ${\color{black} It is straightforward that $\rho_1=\ket{u_1}\bra{u_1}$ and $\rho_2=\ket{u_2}\bra{u_2}$ are distinct invariant states of the auxiliary map $\mathcal{L}$, thus $\mathcal{L}$ can not be ergodic. By Theorem \ref{adap2}, it has
two invariant subspaces $S_1\neq\{0\}$ and $S_2\neq\{0\}$ such that $S_1\cap S_2=\{0\}.$ Since the coin has dimension 2, two distinct pure invariant states must correspond to orthogonal rank-1 projections. Indeed, if $\rho_1$ and $\rho_2$ are both invariant and $\ket{u_1},\ket{u_2}$ are linearly independent, the only way for $\mathcal{L}$ to have both as fixed points is if they are orthogonal. Otherwise, any operator that fixes both $\rho_1$ and $\rho_2$ would fix all operators, forcing $\mathcal{L}$ to be the trivial identity map.}  We assume that $|l_1|=|r_1|$ and $|l_2|=|r_2|,$ then for $\rho=\ket{u}\bra{u},$ where $\ket{u}$ is some common eigenvector of $L,B,R,$ we have $p_{j,j+1;\rho}(n)=p_{j,j-1;\rho}(n)=w$ for all $j\in\mathbb{Z}$ and all $n\geq 0$ for some constant $w$ and therefore, in this case, we have a classical homogeneous walk which is recurrent. If $(L,B,R)$ was not recurrent, then we would be in third situation of proposition \ref{splitprop}, with $$\mathfrak{h}=\mathcal{Y}\oplus\mathcal{X},\quad \text{where}\quad  \text{dim}(\mathcal{Y})\geq 1,  \;\;\text{ and }\;\; \text{dim}(\mathcal{X})=2,
$$
a contradiction. Therefore, we are in the first situation of proposition \ref{splitprop} and the walk is recurrent.

Analogously, if $|l_1|\neq|r_1|$ and $|l_2|\neq|r_2|,$ then for $\rho_1=\ket{u_1}\bra{u_1},$ and $\rho_2=\ket{u_2}\bra{u_2},$ where $\ket{u_1},\ket{u_2}$ are orthonormal common eigenvectors of $L,B,R,$ we have $p_{j,j+1;\rho}(n)< p_{j,j-1;\rho}(n)$ (or $>$) for all $j\in\mathbb{Z}$ and all $n\geq 0$ and therefore in this case we have a classical homogeneous walk which is transient for both $\rho_1,\rho_2$. Since the walk is transient for two distinct eigenvectors, the walk is transient by Corollary \ref{cor2}.

For the last case, $|l_i|=|r_i|$ and $|l_j|\neq|r_j|,$ $(L,B,R)$ is $\rho_i$-recurrent and $\rho_j$-transient, where $\rho_i=\ket{u_i}\bra{u_i}$ and $\rho_j=\ket{u_j}\bra{u_j}.$ This situation is a particular case of the last item in Proposition \ref{splitprop}.

\end{proof}
\end{teo}

\subsection{General Recurrence Criterion for HOQWs Induced by Finite Dimensional Coins on the Line}\label{sec4}

Following the approach of \cite[Section 4]{carboneCLT} and the references therein, we can obtain for a coin $(L,R)$ a decomposition
\begin{equation*}\label{h-def}
\mathfrak{h}=\mathfrak{R}\oplus \mathfrak{T},\qquad
\mathfrak{R}=\sup\{\mbox{supp}(\omega) \;|\;\omega \mbox{ is an invariant state for } \mathcal{L}\}.
\end{equation*}

Since $\mathfrak{h}$ is finite-dimensional, any minimal enclosure is contained in $\mathfrak{R},$ and is the support of a unique extremal invariant state. Moreover, we have a unique decomposition of $\mathfrak{R}$ of the form
\begin{equation}\label{R_grande}
\mathfrak{R}=\bigoplus_{\alpha\in A} \mathfrak{X}_\alpha,
\end{equation}
where $\{\mathfrak{X}_\alpha\}_\alpha\in A$ is a finite set of mutually orthogonal enclosures and every $\mathfrak{X}_\alpha$ is minimal in the set of enclosures satisfying 
$$
\mbox{ for any minimal enclosure }\mathcal{Y}\mbox{ either } \mathcal{Y}\perp\mathfrak{X}_\alpha\mbox{ or }
\mathcal{Y}\subset\mathfrak{X}_\alpha.
$$

Further, every $\mathfrak{X}_\alpha$ is either a minimal enclosure or admits a decomposition (not unique) as the sum of mutually orthogonal isomorphic minimal enclosures $\{\mathcal{Y}_{\alpha,\beta},\;\beta\in I_\alpha\}:$
$$
\mathfrak{X}_\alpha=\bigoplus_{\beta\in I_\alpha}\mathcal{Y}_{\alpha,\beta}.
$$
Given a minimal enclosure $\mathcal{Y}_\beta,$ let $\tau_\beta$ be the unique invariant state with support on $\mathcal{Y}_\beta.$ {\color{black}We introduce the parameter
\[
m_\beta = \Tr\left(L\tau_\beta L^*\right) - \Tr\left(R\tau_\beta R^*\right),
\]
where $\tau_\beta$ is the unique invariant state supported on the minimal enclosure $\mathcal{Y}_\beta$.

\begin{lema}\cite[Lemma 4.2]{carboneCLT}\label{lema4.2}
The parameter $m_\beta$ is independent of the particular minimal enclosure $\mathcal{Y}_\beta$ in $\mathfrak{X}_\alpha$.
\end{lema}

Consequently, $m_\beta$ is a characteristic parameter of the component $\mathfrak{X}_\alpha$ itself, and we may denote it by $m_\alpha$ when referring to the whole component.} Therefore, we can combine Proposition \ref{splitprop}, Theorem \ref{teo_lazy} and the discussion above to state the property

\begin{prop}\label{propms}
Consider a coin $(L,R).$ We have a decomposition
\begin{equation}\label{h_prop}
\mathfrak{h}=\left(\bigoplus_{\alpha\in A}\mathcal{Y}_\alpha\right)\oplus\mathcal{X},
\end{equation}
where $A$ is a finite set, $\mathcal{Y}_\alpha$ is a subspace of $\mathfrak{h}$ being the support of exactly one invariant state $\tau_\alpha$ for $\mathcal{L},$ $\mathcal{X}$ contains no enclosure, and
\begin{itemize}
  \item $\textmd{Tr}\left(L\tau_\alpha L^*\right)=\dfrac{1}{2}\Rightarrow$ the walk is $\tau$-recurrent for all $\tau\in \mathcal{D}(\mathcal{Y}_\alpha)$;

  \bigskip
  
  \item $\textmd{Tr}\left(L\tau_\alpha L^*\right)\neq\dfrac{1}{2}\Rightarrow$ the walk is $\tau$-transient for all $\tau\in \mathcal{D}(\mathcal{Y}_\alpha).$
\end{itemize}
\end{prop}

\medskip

Once the decomposition in Equation \eqref{h_prop} is obtained, we can say that $\mathcal{Y}_\alpha$ is either recurrent or transient.

The following result is due to \cite{carboneCLT} and will be crucial to our next criterion. To state it, consider a coin $(L,R),$ and let $\mathcal{Y}$ be an enclosure for $\mathcal{L}.$ Denoting $p_{\mathcal{Y}}$ the orthogonal projection onto $\mathcal{Y},$ we can define the associated absorption operator (see \cite{carbone-absorption}):
$$
A(\mathcal{Y}):=\lim_{n\rightarrow\infty}\mathcal{L}^{*n}(p_{\mathcal{Y}}).
$$
\begin{corollary}\cite[Corollary 4.5]{carboneCLT}\label{corlaw} Consider a coin $(L,R)$ with the decomposition of $\mathfrak{h}$ given in Equation \eqref{h_prop}, and let
$a_\alpha(\rho)=\mathbb{E}_\rho\left[\Tr\left(A(\mathcal{Y}_\alpha)\rho_0\right)\right].$
Then we have the convergence in law
\begin{equation}\label{large-numbers}
\frac{X_n-X_0}{n}\stackrel{n\rightarrow\infty}{\longrightarrow}
\sum_{\alpha\in A} a_\alpha(\rho)\delta_{m_\alpha},
\end{equation}
where $\delta_{m_\alpha}$ is the Dirac measure concentrated in $m_\alpha.$
\end{corollary}

We remark that the convergence in law brought by Corollary \ref{corlaw} can be upgraded to a convergence a.s., as demonstrated in \cite[Theorem 3.5.2]{girotti-thesis}.

Let us start an HOQW induced by a coin $(L,R).$ There are two possibilities concerning recurrence when  
$
\text{supp}(\rho_0)\not\perp \mathcal{X}:$

\bigskip

\textbf{1. $(L,R)$ is recurrent for $\mathcal{Y}_\alpha$ for some $\alpha$ and supp$\left(\mathcal{L}^n(\rho_0)\right)\not\perp\mathcal{Y}_\alpha$ for some $n\in\mathbb{N}$:} in this case, the walk is $\rho_0$-recurrent. Indeed, concerning the quantum trajectories $(\rho_n)_n$, we could refresh the walk at some instant $n$ at some vertex $\ket{X_n},$ which is $\rho_n$-recurrent, since supp$\left(\mathcal{L}^n(\rho_0)\right)\not\perp\mathcal{Y}_\alpha$ for a sufficiently large $n$. If there was some density $\sigma$ with supp$(\sigma)\not\perp\mathcal{X}$ and $\sigma$-transient, there would be some enclosure not orthogonal to $\mathcal{X},$ which is a contradiction by Proposition \ref{splitprop}. The conclusion is held by the homogeneity of the coin, since, in this case, the walk either is $\rho_s$-recurrent for every $s=0,1,2,\ldots$ or it is $\rho_s$-transient for every $s=0,1,2,\ldots$, analogously to the proof of Proposition \ref{splitprop}.

\bigskip

\textbf{2. $(L,R)$ is transient for all $\mathcal{Y}_\alpha$ satisfying supp$\left(\mathcal{L}^n(\rho_0)\right)\not\perp\mathcal{Y}_\alpha$ for every $n\in\mathbb{N}$:}  the walk is $\sigma$-transient for all $\sigma\in\mathcal{D}(\mathcal{X})$. Indeed, in this case, starting the walk at vertex $X_0=0,$ $X_n/n$ converges (when $n\rightarrow \infty$) to 
$$m=\sum_{\alpha\in A} a_\alpha(\rho)\delta_{m_\alpha}=\sum_{\alpha\in A} a_\alpha(\rho),$$
as given by Equation \eqref{large-numbers}, where the second equality follows from $m_\alpha\neq 0,$ since $(L,R)$ is $\mathcal{Y}_\alpha$ transient.
Thus, $X_n$ converges either to $-\infty$ or $\infty,$ in probability, imposing that the walk is $\rho$-transient and thus $(L,R)$ is transient on $\mathcal{D}(\mathcal{X}).$

Therefore, we have shown the following recurrence criterion:

\begin{teo}[A Generalized Criteria]\label{TeoGeneralCriteria} Consider a coin $(L,R)$ of dimension $d.$ Consider the decomposition of $\mathfrak{h}$ as given in Equation \eqref{h_prop}, and $\tau_1,\ldots,\tau_m$ the respectively invariant states of $\mathcal{L}$ in $\mathcal{Y}_{\alpha},\alpha=1,\ldots,m.$

\medskip

$(1)\quad\Tr(L\tau_j L^*)\neq \dfrac{1}{2}\;\forall j\in\{1,\ldots m\}\Rightarrow (L,R) \mbox{ is transient;}$ 

\medskip

$(2)\quad\Tr(L\tau_{j_k} L^*)= \dfrac{1}{2}\;$ for a set of indices $j_k\in J \subset\{1 ,\ldots,m\}$ and $\Tr(L\tau_{j_l}L^*)\neq \dfrac{1}{2}\;$ for the remaining indices $j_l\in J^C,$ and $\sigma$ is a density such that supp$(\sigma)\not\perp\mathcal{X}.$

(2.1) supp$\left(\mathcal{L}^n(\sigma)\right)\not\perp\mathcal{Y}_\alpha$ for some $\alpha\in J$ and some $n\in\mathbb{N},$ then $(L,R)$ is
$$\begin{cases}\rho
\text{-transient for } \rho\in \mathcal{D}_1:=\mathcal{D}(\oplus_{k\in J^C}\mathcal{Y}_k)\\
\rho\text{-recurrent for } \rho\in \mathcal{D}_1^C.
\end{cases}$$

\medskip

(2.2) supp$\left(\mathcal{L}^n(\rho)\right)\perp\mathcal{Y}_\alpha$ for all $\alpha\in J$ and all $n\in\mathbb{N},$ then $(L,R)$ is
$$\begin{cases}
\rho\text{-transient for } \rho\in \mathcal{D}_2:=\mathcal{D}(\oplus_{k\in J^C}\mathcal{Y}_k\oplus \mathcal{X})\\
\rho\text{-recurrent for } \mathcal{D}_2^C
\end{cases}$$

\medskip

$(3)\quad\Tr(L\tau_{\alpha} L^*)= \dfrac{1}{2}\;$ for every $\alpha\Rightarrow$ the walk is recurrent.

\end{teo}

\medskip

\begin{REM}
It is worth observing that $\mathcal{D}_2$ in Theorem \ref{TeoGeneralCriteria} actually arises from an enclosure due to the orthogonality between supp$\left(\mathcal{L}^n(\rho)\right)$ and $\mathcal{Y}_\alpha$; otherwise, it would contradict Proposition \ref{splitprop}.
\end{REM}





\subsection{A Generalized Recurrence Criterion for HOQWs on the Grid}.

We have already presented some recurrence criteria for discrete-time and continuous-time OQWs on the set of vertices $\mathbb{Z}^2$, constrained by a minimal enclosure. These criteria are interesting tools for understanding the long-term behavior of OQWs. Building on these consequences, we aim to derive a stronger recurrence criterion for discrete-time HOQWs, allowing for a more comprehensive analysis of their statistics. 

At this point we will allow the auxiliary map of a coin $(D)$ to have more than one invariant state, thus we consider the decomposition of $\mathfrak{h}$ given on Section \ref{sec4}. Therefore, given a minimal enclosure $\mathcal{Y}_\alpha,$ let $\tau_\alpha$ be the unique invariant state with support on $\mathcal{Y}_\alpha.$ Consider the parameter
$$\tilde{m}_\alpha=\begin{bmatrix}
m_{1,\alpha}\\m_{2,\alpha}
\end{bmatrix},\quad
m_{1,\alpha}=\Tr\left(D_1\tau_\alpha D_1^*\right)-\Tr\left(D_3\tau_\alpha D_3^*\right),\quad
m_{2,\alpha}=\Tr\left(D_2\tau_\alpha D_2^*\right)-\Tr\left(D_4\tau_\alpha D_4^*\right).
$$
{\color{black}As stated in} Lemma \ref{lema4.2}, the value $\tilde m_\alpha$ is independent of the particular minimal enclosure $\mathcal{Y}_\alpha$ in $\mathfrak{X}_\alpha.$  Therefore, we can combine Theorem \ref{OQWcriterion} and results concerning CTOQWs above to state the next property, which is a continuous-time version on the grid of Proposition \ref{propms}.

\begin{prop}
Consider a coin $(D).$ We have a decomposition
\begin{equation}\label{h_propCT}
\mathfrak{h}=\left(\bigoplus_{\alpha\in A}\mathcal{Y}_\alpha\right)\oplus\mathcal{X},
\end{equation}
where $A$ is a finite set, $\mathcal{Y}_\alpha$ is a subspace of $\mathfrak{h}$ being the support of exactly one invariant state $\tau_\alpha$ for $\Phi,$ $\mathcal{X}$ contains no enclosure, and
\begin{itemize}
  \item $\tilde{m}_\alpha=0\Rightarrow$ the walk is $\tau_\alpha$-recurrent for all $\tau\in \mathcal{D}(\mathcal{Y}_\alpha)$;

  \bigskip
  
  \item $\tilde{m}_\alpha \neq0\Rightarrow$ the walk is $\tau_\alpha$-transient for all $\tau\in \mathcal{D}(\mathcal{Y}_\alpha).$
\end{itemize}
\end{prop}
\begin{proof}
Fixed a subspace $\mathcal{Y}_\alpha$ which is the support of exactly one invariant state $\tau_\alpha\in\mathcal{D}(\mathfrak{h}),$ we assume that $\tilde{m}_\alpha=0.$ Then $(D)$ is $\rho$-recurrent for every $\rho$ by Theorem \ref{TeoGauss2}. If $\tilde m_\alpha\neq 0,$ then $(D)$ is transient by Theorem \ref{tran}.
\end{proof}

With the decomposition in Equation \eqref{h_propCT} established, we can say that $\mathcal{Y}_\alpha$ is either recurrent or transient. It remains to understand what happens when a quantum state has support outside of $\oplus_{\alpha}\mathcal{Y}_\alpha.$

Analogously to the one dimensional case, our recurrence criterion is:

\begin{teo}[A Generalized Criterion for coins $(D)$]\label{TeoGeneralCriteria2} Consider a coin $(D)$ of dimension $d<\infty.$ Consider the decomposition of $\mathfrak{h}$ as given in Equation \eqref{h_propCT}, and $\tau_1,\ldots,\tau_w$ the respectively invariant states of $\Phi$ in $\mathcal{Y}_{\alpha},\;\alpha=1,\ldots,w.$

\medskip

$(1)\quad \tilde m_\alpha\neq 0\;\forall j\in\{1,\ldots w\}\Rightarrow (D) \mbox{ is transient;}$ 

\medskip

$(2)\quad \tilde m_\alpha= 0\;$ for a set of indices $j_k\in J \subset\{1 ,\ldots,w\}$ and $ \tilde m_\alpha\neq 0\;$ for the remaining indices $j_l\in J^C,$ and $\sigma$ is a density such that supp$(\sigma)\not\perp\mathcal{X}.$

(2.1) supp$\left(\Phi^n(\rho)\right)\not\perp\mathcal{Y}_\alpha$ for some $\alpha\in J$ and some $n\in\mathbb{N},$ then $(D)$ is
$$\begin{cases}\rho
\text{-transient for } \rho\in \mathcal{D}_1:=\mathcal{D}(\oplus_{k\in J^C}\mathcal{Y}_k)\\
\rho\text{-recurrent for } \rho\in \mathcal{D}_1^C.
\end{cases}$$

\medskip

(2.2) supp$\left(\Phi^n(\rho)\right)\perp\mathcal{Y}_\alpha$ for all $\alpha\in J$ and all $n\in\mathbb{N},$ then $(D)$ is
$$\begin{cases}
\rho\text{-transient for } \rho\in \mathcal{D}_2:=\mathcal{D}(\oplus_{k\in J^C}\mathcal{Y}_k\oplus \mathcal{X})\\
\rho\text{-recurrent for } \mathcal{D}_2^C
\end{cases}$$

\medskip

$(3)\quad \tilde m_\alpha= 0\;$ for every $\alpha\Rightarrow$ the walk is recurrent.

\end{teo}

{\color{black}For quick reference, Table~\ref{tab:notation} gathers the notation for the density sets and operator spaces employed in the paper.

\begin{table}[h!]
\centering
\begin{tabular}{|c|l|}
\hline
\textbf{Notation} & \textbf{Description} \\
\hline
$\mathfrak{h}$ & Internal space \\
\hline
$\mathcal{D}(\mathfrak{h})$ & Set of density operators on the internal space $\mathfrak{h}$ \\
\hline
$\mathcal{D}_i(\mathfrak{h})$ & Set of density operators  on the internal space $\mathfrak{h}_i$ \\
\hline
$\mathcal{S}(\mathcal{H}\otimes\mathcal{K})$ & Set of density operators on $\mathcal{H}\otimes\mathcal{K}$ of diagonal form with respect to position \\
\hline
$\mathcal{D}(\mathcal{Y})$ & Set of density operators supported in a subspace $\mathcal{Y}\subset\mathfrak{h}$ \\
\hline
$\mathcal{D}_1$, $\mathcal{D}_2$ & Sets of transient densities in Theorems \ref{TeoGeneralCriteria} and \ref{TeoGeneralCriteria2} \\
\hline
$\mathcal{I}_1(\mathcal{H})$ & Trace-class operators on $\mathcal{H}$ \\
\hline
$\mathcal{B}(\mathcal{W})$ & Bounded linear operators on a Hilbert space $\mathcal{W}$ \\
\hline
\end{tabular}
\caption{Notation for  operator sets and density spaces.}
\label{tab:notation}
\end{table}
}

{\color{black}
Having established rigorous recurrence criteria for HOQWs and CTOQWs on the line and on the grid, which classify the behavior based on the decomposition of the internal space $\mathfrak{h}$ and the drift parameters $\tilde{m}_\alpha$, we now illustrate these concepts with explicit numerical matrices. The following examples are also designed to demonstrate how the decomposition into enclosures are performed in practice.
}
\section{Examples}

\subsection{1D}
\begin{ex}\label{ex1}Our first examples will consider Lazy HOQWs of dimension 2.

\begin{enumerate}
\item Let $m,n>0$ and assume that $b:=1-m^2-n^2\in(0,1).$ Then the coin $(L,B,R)$ induced by
$$
L=\begin{bmatrix}
0&0\\
0&-n
\end{bmatrix},\qquad
B=\begin{bmatrix}
-\sqrt{b}&m\\
m&\sqrt{b}
\end{bmatrix},\qquad
R=\begin{bmatrix}
n&0\\
0&0
\end{bmatrix}
$$
is well-defined.

The only invariant state for the auxiliary map is $\rho_\infty =I/2.$ Therefore,
$$
\Tr(L\rho_\infty L^*)=\frac{n^2}{2}=\Tr(R\rho_\infty R^*)\Rightarrow\;(L,B,R) \mbox{ is recurrent,}
$$
by Item (1) of Theorem \ref{LazyDim2}.

{\color{black}Note that for $m=0$, the operators $L,B,R$ have the common and linearly independent eigenvectors $[1,0]^T$ and $[0,1]^T$. According to item (2.2) of Theorem \ref{LazyDim2}, with $|l_1|=0$, $|l_2|=n$, $|r_1|=n$, $|r_2|=0$, the walk is transient.}

\begin{flushright}
$\Diamond$
\end{flushright}

\medskip

\item Let $x_1,x_2\in \mathbb{C},$
$$
|x_1|\leq\sqrt{\frac{2}{3}},\quad |x_2|\leq\sqrt{\frac{3}{4}},$$
then consider the coin $(L,B,R),$ where
$$
L=\begin{bmatrix}
\frac{1}{\sqrt{3}}&0\\
0&\frac{1}{2}
\end{bmatrix},\qquad
B=\begin{bmatrix}
\sqrt{\frac{2}{3}-x_1^2}&0\\
0&\sqrt{\frac{3}{4}-x_2^2}
\end{bmatrix},\qquad
R=\begin{bmatrix}
x_1&0\\
0&x_2
\end{bmatrix}.
$$
We are in Item (2) of Theorem \ref{LazyDim2}, thus we conclude that

\begin{eqnarray}
\nonumber (1)\quad |x_1|=\frac{1}{\sqrt{3}}\mbox{ and }|x_2|=\frac{1}{2}&\Rightarrow& (L,B,R)\mbox{ is recurrent,} \\
\nonumber &\\
\nonumber (2)\quad |x_1|\neq\frac{1}{\sqrt{3}}\mbox{ and }|x_2|\neq\frac{1}{2}&\Rightarrow& (L,B,R)\mbox{ is transient,} \\
\nonumber &\\
\nonumber (3)\quad |x_1|=\frac{1}{\sqrt{3}}\mbox{ and }|x_2|\neq\frac{1}{2}&\Rightarrow& (L,B,R)\mbox{ is transient with respect to }\sigma=\begin{bmatrix}
0&0\\0&1
\end{bmatrix}\mbox{ and it is} \\
\nonumber   & & \mbox{recurrent with respect to all densities but }\sigma,\\
\nonumber (4)\quad |x_1|\neq\frac{1}{\sqrt{3}}\mbox{ and }|x_2|=\frac{1}{2}&\Rightarrow& (L,B,R)\mbox{ is transient with respect to }\sigma=\begin{bmatrix}
1&0\\0&0
\end{bmatrix}\mbox{ and it is} \\
\nonumber   & & \mbox{recurrent with respect to all densities but }\sigma.
\end{eqnarray}
\end{enumerate}
\qee
\end{ex}

\begin{ex}
Consider the Lazy HOQW induced by the following coin of dimension 3: $(L,B,R),$ where
$$
L=\frac{1}{8}\begin{bmatrix}
2(1+\sqrt{2})&0&2(1-\sqrt{2})\\
0&\sqrt{31}&0\\
2(1-\sqrt{2})&0&2(1+\sqrt{2})
\end{bmatrix},\quad
B=\frac{1}{16}\begin{bmatrix}
\sqrt{30}&2i&\sqrt{30}\\
2&0&2\\
\sqrt{30}&-2i&\sqrt{30}
\end{bmatrix},\quad
R=\frac{1}{8}\begin{bmatrix}
2(1-\sqrt{2})&0&2(1+\sqrt{2})\\
0&\sqrt{31}&0\\
2(1+\sqrt{2})&0&2(1-\sqrt{2})
\end{bmatrix}.
$$

There is a unique invariant state $\rho_\infty$ for the auxiliary map given by
$$
\rho_\infty=\frac{1}{2}
\begin{bmatrix}
1&0&-1\\
0&0&0\\
-1&0&1
\end{bmatrix},
$$
and $\Tr\left(L\rho_\infty L^*\right)=1/2=\Tr\left(R\rho_\infty R^*\right).$ Therefore, this coin is recurrent by Theorem \ref{teo_lazy}.
\qee
\end{ex}

Since the invariant state also depends on the choice of $B,$ it is natural to ask if for a given transient coin $(L,B,R),$ there is another $B'$ that makes $(L,B',R)$ to be recurrent.

\begin{ex}\label{ex3} 
Firstly, let
$$
L=\frac{1}{7}\begin{bmatrix}
1&0\\
0&2
\end{bmatrix},\qquad
B=\frac{1}{7}\begin{bmatrix}
-1&1\\
2&0
\end{bmatrix},\qquad
R=\frac{1}{7}   \begin{bmatrix}
1&1\\
0&1
\end{bmatrix}.
$$

Again, there is a unique invariant state $\rho_\infty$ for the auxiliary map $\mathcal{L}$ given by
$$\rho_\infty=\frac{1}{3}\begin{bmatrix}
1&0\\
0&2
\end{bmatrix}.$$ In this case, $\Tr\left(L\rho_\infty L^*\right)=3/7$ and $\Tr\left(R\rho_\infty R^*\right)=5/21,$ thus the coin $(L,B,R)$ is transient by item (1) of Theorem \ref{LazyDim2}. However, if we take
$$
B'=\frac{1}{\sqrt{35}}\begin{bmatrix}
5&-1\\
0&2
\end{bmatrix},
$$
then the unique invariant state for the auxiliary map $\mathcal{L}'$ is
$$
\rho'_\infty=
\begin{bmatrix}
1&0\\
0&0
\end{bmatrix},
$$
and
$\Tr\left(L\rho_\infty L^*\right)=1/\sqrt{7}=\Tr\left(R\rho_\infty R^*\right).$ Thus the coin $(L,B',R)$ is recurrent, also by item (1) of Theorem \ref{LazyDim2}
\end{ex}
\qee

{\color{black}We remark that this choice was possible because $L$ and $R$ share exactly one common eigenvector, namely $\ket{v}=[1,0]^T$, and satisfy $L\ket{v}=R\ket{v}$. Under these conditions, $\Tr(L\rho L^*) = \Tr(R\rho R^*)$ holds for $\rho = \ket{v}\bra{v}$. Hence, recurrence could be achieved by finding a matrix $B'$ such that:
\begin{enumerate}
    \item $\ket{v}$ is also an eigenvector of $B',$ thus $\rho$ becomes the unique invariant state for $\mathcal{L}'$;
    \item the normalization condition $L^*L + B'^*B' + R^*R = I$ is satisfied.
\end{enumerate}
The explicit matrix $B'$ presented in the example was obtained by solving these constraints. It is worth noting that finding such $B'$ for more general $L$ and $R$ is considerably more difficult. Several attempts were made using mathematical software (Maple) with various choices of $L$ and $R$ having simple entries; however, the resulting matrices $B'$ that yielded recurrence often had very lengthy expressions, making them impractical to display in this paper.

Another natural continuation of this work would be to investigate under what conditions on $L$ and $R$ there exists a matrix $B'$ such that $(L,B',R)$ is recurrent, and to develop methods for explicitly constructing such $B'$. Given fixed $L,R$, one would like to determine whether the set of admissible $B$ (satisfying $L^*L+B^*B+R^*R=I$) contains recurrent elements, and to describe them explicitly. As seen in this work, this problem is closely tied to the invariant state $\rho_\infty$ of $\mathcal{L}$ and the condition $\Tr(L\rho_\infty L^*) = \Tr(R\rho_\infty R^*)$.
}

\begin{ex} Let us consider the coin given in \cite[Section 6.2]{carboneCLT}:
$$R=
\begin{bmatrix}
\sqrt{\frac{3}{8}}&0&0&0\\
-\sqrt{\frac{p_1}{2}}&\frac{1}{\sqrt{2}}&0&0\\
-\sqrt{\frac{p_2}{2}}&0&\frac{1}{\sqrt{2}}&0\\
\sqrt{\frac{2p_3}{3}}&0&0&\frac{1}{\sqrt{3}}
\end{bmatrix},\qquad
L=
\begin{bmatrix}
\frac{1}{2\sqrt{2}}&0&0&0\\
\sqrt{\frac{p_1}{2}}&\frac{1}{\sqrt{2}}&0&0\\
\sqrt{\frac{p_2}{2}}&0&\frac{1}{\sqrt{2}}&0\\
-\sqrt{\frac{p_3}{3}}&0&0&\frac{\sqrt{2}}{\sqrt{3}}
\end{bmatrix},
$$
where $p_1,\;p_2,p_3\geq 0$ and $p_1/2+p_2/2+p_3/2=1/2.$

The invariant states of the auxiliary map are of the form $\rho_\infty=\sigma_1+\sigma_2,$ where
$$
\sigma_1=
\omega_1\begin{bmatrix}
0&0&0&0\\
0&a&b&0\\
0&b^*&1-a&0\\
0&0&0&0
\end{bmatrix},\quad
\sigma_2=
\omega_2\begin{bmatrix}
0&0&0&0\\
0&0&0&0\\
0&0&0&0\\
0&0&0&1
\end{bmatrix},\quad \omega_i,\omega_j\geq 0,\;\;\omega_1+\omega_2=1.
$$

We have
$$
m_1=\Tr(L\sigma_1 L^*)=\frac{1}{2} ,\qquad
m_2=\Tr(L\sigma_2 L^*)=\frac{2}{3} \neq \frac{1}{2}.
$$

By item (2) of Theorem \ref{TeoGeneralCriteria}, the walk is transient for
$$
\rho=
\begin{bmatrix}
0&0&0&0\\
0&0&0&0\\
0&0&0&0\\
0&0&0&1
\end{bmatrix}
$$
and recurrent for all the remaining density operators on $\mathbb{C}^4$, for any appropriate choices of $p_k,\;k=1,2,3.$

\end{ex}

\begin{ex} Consider an HOQW induced by the coin $(L,R),$ where
$$L=\begin{bmatrix}
\frac{\sqrt{5}}{5}&0&-\frac{\sqrt{5}}{5}\\
0&\frac{2\sqrt{5}}{5}&\frac{\sqrt{5}}{10}\\
0&0&\frac{1}{2}
\end{bmatrix},\qquad
R=\begin{bmatrix}
\frac{2\sqrt{5}}{5}&0&\frac{\sqrt{5}}{10}\\
0&\frac{\sqrt{5}}{5}&-\frac{\sqrt{5}}{5}\\
0&0&\frac{1}{2}
\end{bmatrix}.
$$

The invariant states of the auxiliary map are of the form $\rho_\infty=\sigma_1+\sigma_2,$ where
$$
\sigma_1=
\omega_1\begin{bmatrix}
1&0&0\\
0&0&0\\
0&0&0
\end{bmatrix},\quad
\sigma_2=
\omega_2\begin{bmatrix}
0&0&0\\
0&1&0\\
0&0&0
\end{bmatrix},\quad \omega_i,\omega_j\geq 0,\;\;\omega_1+\omega_2=1.
$$

We have
$$
m_1=\Tr(L\sigma_1 L^*)=\frac{\sqrt{5}}{5}\neq\frac{1}{2} ,\qquad
m_2=\Tr(L\sigma_2 L^*)=\frac{2\sqrt{5}}{5}\neq\frac{1}{2}.
$$

By item (1) of Theorem \ref{TeoGeneralCriteria}, the walk is transient.

\end{ex}

\begin{ex}\label{ex4} Let us consider the coin $(L,R)$ of dimension 4 given by
$$R=
\begin{bmatrix}
\frac{\sqrt{2}}{2}&-\frac{\sqrt{5}}{4}&0&\frac{1}{4}\\
0&\frac{\sqrt{2}}{4}&0&0\\
0&0&\frac{\sqrt{2}}{2}&0\\
0&0&0&\frac{\sqrt{6}}{4}
\end{bmatrix},\qquad
L=
\begin{bmatrix}
\frac{\sqrt{2}}{2}&\frac{\sqrt{5}}{4}&0&-\frac{1}{4}\\
0&\frac{1}{2}&0&\frac{\sqrt{5}}{4}\\
0&0&-\frac{\sqrt{2}}{2}&0\\
0&0&0&\frac{\sqrt{3}}{4}
\end{bmatrix},
$$

The invariant states of the auxiliary map are of the form $\rho_\infty=\sigma_1+\sigma_2,$ where
$$
\sigma_1=
\omega_1\begin{bmatrix}
1&0&0&0\\
0&0&0&0\\
0&0&0&0\\
0&0&0&0
\end{bmatrix},\quad
\sigma_2=
\omega_2\begin{bmatrix}
0&0&0&0\\
0&0&0&0\\
0&0&1&0\\
0&0&0&0
\end{bmatrix},\quad \omega_i,\omega_j\geq 0,\;\;\omega_1+\omega_2=1.
$$

We have
$$
m_1=\Tr(L\sigma_1 L^*)=\frac{1}{2} ,\qquad
m_2=\Tr(L\sigma_2 L^*)=\frac{1}{2}.
$$

By item (3) of Theorem \ref{TeoGeneralCriteria}, the walk is recurrent.
\end{ex}

\subsection{2D}

\begin{ex}
We start with a CTOQW induced by a coin $(A,H),$ 
$$A_1 = \begin{bmatrix}
3 & -1 \\
0 & 0
\end{bmatrix}, \quad
A_2 = \begin{bmatrix}
1 & -2 \\
2i & 0
\end{bmatrix}, \quad
A_3 = \begin{bmatrix}
1 & 1 \\
-2 & 2
\end{bmatrix}, \quad
A_4 = \begin{bmatrix}
-2i & i \\
0 & 2
\end{bmatrix}, \quad
H = \begin{bmatrix}
-1 & h \\
\overline{h} & 2
\end{bmatrix},\;h\in\mathbb{C}.
$$

Operator $\mathbb{L}$ has a unique stationary state $\rho_\infty,$ where
$$
\rho_\infty=\begin{bmatrix}
\rho_1&\rho_2\\ 
\rho_2&1-\rho_1
\end{bmatrix},
\quad
\rho_1=\dfrac{19(\text{Re}\,h)^2 - 8(\text{Re}\,h)(\text{Im}\,h) + 15(\text{Im}\,h)^2 - 54\,\text{Re}\,h + 100\,\text{Im}\,h + 830}{\alpha},
$$
$$
\rho_2=\dfrac{20i(\text{Re}\,h)(\text{Im}\,h) - 39i(\text{Re}\,h) - 37i(\text{Im}\,h) + 20(\text{Re}\,h)^2 - 45(\text{Re}\,h) + 15(\text{Im}\,h) + 1165 + 309i}{2\alpha},
$$
$\alpha=38(\text{Re}\,h)^2 - 16(\text{Re}\,h)(\text{Im}\,h) + 30(\text{Im}\,h)^2 - 78\,\text{Re}\,h + 50\,\text{Im}\,h + 1945.$ In this case,
$$
m=\frac{2\; \text{Re}\;h-10\; \text{Im}\;h+19}{\alpha}\;\begin{bmatrix}
4\\
1
\end{bmatrix},
$$
Therefore, by Theorem \ref{ct.criterion}, $(A,H)$ is recurrent if and only if 
$$
\text{Re}(h)=5\;\text{Im}(h)-\frac{19}{2}.
$$
\qee\end{ex}

\begin{ex}\label{ex2} Consider an HOQW induced by a coin $(D),$ where
$$D_1=\frac{1}{2\sqrt{30}}\begin{bmatrix}
2&0&-2\\
0&4&1\\
0&0&\sqrt{5}
\end{bmatrix},\;
D_2=\frac{1}{2\sqrt{6}}\begin{bmatrix}
2&0&-2\\
0&4&1\\
0&0&\sqrt{5}
\end{bmatrix},\;
D_3=\frac{1}{2\sqrt{30}}\begin{bmatrix}
4&0&1\\
0&2&-2\\
0&0&\sqrt{5}
\end{bmatrix},\;
D_4=\frac{1}{2\sqrt{6}}\begin{bmatrix}
4&0&1\\
0&2&-2\\
0&0&\sqrt{5}
\end{bmatrix}.
$$

The invariant states of the auxiliary map are of the form $\rho_\infty=\omega_1\sigma_1+\omega_2\sigma_2,$ where
$$
\sigma_1=
\begin{bmatrix}
1&0&0\\
0&0&0\\
0&0&0
\end{bmatrix},\quad
\sigma_2=
\begin{bmatrix}
0&0&0\\
0&1&0\\
0&0&0
\end{bmatrix},\quad \omega_i,\omega_j\geq 0,\;\;\omega_1+\omega_2=1.
$$

We have
$$
\tilde{m}_1=\frac{1}{10}\begin{bmatrix}
-1\\
-5
\end{bmatrix},\qquad
\tilde m_2=
\frac{1}{10}\begin{bmatrix}
1\\
5
\end{bmatrix}.
$$

By item (1) of Theorem \ref{TeoGeneralCriteria2}, the walk is transient.

\end{ex}

\begin{ex}\label{ex8}
Consider the coins $(A,H_1)$ and $(A,H_2),$ where $A_k=D_k$ for $k=1,2,3,4$ as in Example \ref{ex2}, and
$$
H_1=\begin{bmatrix}
0&0&0\\
0&0&0\\
0&0&0
\end{bmatrix},\qquad
H_2=\begin{bmatrix}
0&1&0\\
1&0&0\\
0&0&0
\end{bmatrix}.
$$

It is straightforward that $(A,H_1)$ is transient, since we can join the results of Example \ref{ex2} and Theorem \ref{d.vs.c}. However, it is very interesting to see that $(A,H_2)$ is recurrent. Indeed, 
$$
\rho_\infty=\frac{1}{2}
\begin{bmatrix}
1&0&0\\
0&1&0\\
0&0&0
\end{bmatrix}
$$
is the only stationary state and $m=0,$ thus the coin is recurrent by Theorem \ref{ct.criterion}.
\qee\end{ex}

{\color{black}In Example \ref{ex3}, the operator $H_2$ plays a role analogous to that of $B'$ in Example \ref{ex8}. Given the transition operators $A_k$ (with $k=1,2,3,4$), the choice of $H_2$ alters the stationary state $\rho_\infty$ of the auxiliary map, yielding $m = 0$ and consequently recurrence for the CTOQW. This illustrates how, in both discrete and continuous-time settings, modifying the ``non-jump" component (whether $B$ in the lazy HOQW case or $H$ in the CTOQW case) can tune the drift parameters and induce recurrence, even when the transition operators alone would lead to transience.}

\bmhead{Acknowledgements}
This work was supported by Universidade Federal do Rio Grande do Sul (UFRGS). The author also thanks the reviewers for their comments and suggestions, which significantly improved the quality of the manuscript.

\bmhead{Conflict of interest} The author declare that there is no conflict of interest.

\bmhead{Data Availability} The datasets generated during the current study are available from the corresponding
author on reasonable request.




\end{document}